\documentclass[showpacs,preprint,preprintnumbers,amsmath,amssymb,floatfix]{revtex4-1}
\usepackage{graphicx}
\usepackage{dcolumn}
\usepackage{bm}

\begin{document}

\title{Magnetic Field Effects on Transport Properties of PtSn$_{4}$}

\author{Eundeok Mun$^{1}$, Hyunjin Ko$^{2}$, Gordon J. Miller$^{2}$, German D. Samolyuk$^{1}$\footnote{Current address : Materials Science and Technology Division, Oak Ridge National Laboratory}, Sergey L. Bud'ko$^{1}$, and Paul. C. Canfield$^{1}$}
\affiliation{$^{1}$Ames Laboratory US DOE and Department of Physics and Astronomy, Iowa State University, Ames, IA 50011, USA}%
\affiliation{$^{2}$Ames Laboratory US DOE and Department of Chemistry, Iowa State University, Ames, IA 50011, USA}%

\begin{abstract}

The anisotropic physical properties of single crystals of orthorhombic PtSn$_{4}$ are reported for magnetic fields up to 140\,kOe, applied parallel and perpendicular to the
crystallographic \textbf{b}-axis. The magnetic susceptibility has an approximately temperature independent behavior and reveals an anisotropy between \textbf{ac}-plane and
\textbf{b}-axis. Clear de Haas-van Alphen oscillations in fields as low as 5\,kOe and at temperatures as high as 30\,K were detected in magnetization isotherms. The thermoelectric
power and resistivity of PtSn$_{4}$ show the strong temperature and magnetic field dependencies. A change of the thermoelectric power at $H$ = 140\,kOe is observed as high as $\simeq$
50\,$\mu$V/K. Single crystals of PtSn$_{4}$ exhibit very large transverse magnetoresistance of $\simeq$ 5$\times$10$^{5}$\,\% for the \textbf{ac}-plane and of $\simeq$
1.4$\times$10$^{5}$\,\% for the \textbf{b}-axis resistivity at 1.8\,K and 140\,kOe, as well as pronounced Shubnikov de Haas oscillations. The magnetoresistance of PtSn$_{4}$ appears
to obey Kohler's rule in the temperature and field range measured.  The Hall resistivity shows a linear temperature dependence at high temperatures followed by a sign reversal around
25\,K which is consistent with thermoelectric power measurements. The observed quantum oscillations and band structure calculations indicate that PtSn$_{4}$ has three dimensional
Fermi surfaces.
\end{abstract}


\maketitle

\section{Introduction}

In the Pt-Sn binary system, five stoichiometric binary compounds Pt$_{3}$Sn, PtSn, Pt$_{2}$Sn$_{3}$, PtSn$_{2}$, and PtSn$_{4}$ have been reported \cite{Okamoto}, whereas Pt$_{3}$Sn
and PtSn melt congruently, Pt$_{2}$Sn$_{3}$, PtSn$_{2}$, and PtSn$_{4}$ decompose peritectically. The Pt-Sn system has been proposed for a wide range of application such as soldering,
dentistry, and jewelry \cite{Kuhmann, Kempf, Biggs}. Interestingly, it is observed that the intermetallic phase of PtSn$_{4}$ is formed during soldering at the interface of
metallization and solder; the dissolution of Pt in SnPb (60/40) solder was found to be linear and the rate of dissolution at 260\,$^{\texttt{o}}$C was 24\,nm/min \cite{Kuhmann}. Only
few studies of physical properties on the Sn-rich side have been performed on polycrystalline samples \cite{Kanekar, Kubiak, Kunnen}. Although PtSn$_{4}$ is a deeply peritectic
compound, on the Sn-rich side (95-99\,\% Sn) it is relatively easy to grow single crystals of PtSn$_{4}$ out of a Sn-rich solution.

\begin{figure}
\centering
\includegraphics[width=1\linewidth]{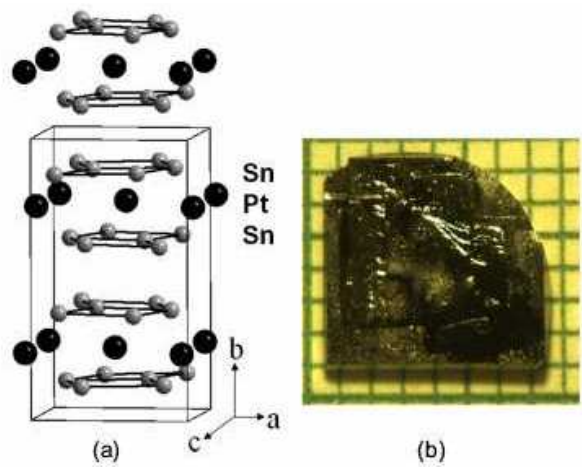}
\caption{(a) Crystal structure of PtSn$_{4}$; along \textbf{b}-axis showing Sn-Pt-Sn layers. (b) Photo of a representative single crystal over a millimeter grid. The \textbf{b}-axis
is perpendicular to the plate.}
\label{structure}%
\end{figure}

PtSn$_{4}$ compound crystallizes in an orthorhombic structure (PtSn$_{4}$-type, Ccca (No.68), $z$ = 4) with lattice constants $a$ = 6.418\,\AA, $b$ = 11.366\,\AA, and $c$ =
6.384\,\AA\, \cite{Kunnen}. This structure can be visualized as consisting of layers which are a stacking of Sn-Pt-Sn layers along the crystallographic \textbf{b}-axis, as shown in
Fig. \ref{structure} (a). As could be expected from the crystal structure with its pseudo tetragonal symmetry, $a\,\simeq\,c\,\neq\,b$, the electrical resistivity and magnetization
measurements indicate an axial anisotropy between \textbf{ac}-plane and \textbf{b}-axis with negligible \textbf{ac}-in-plane anisotropy.

Here we report the physical properties of single crystals of PtSn$_{4}$. The observed electronic and magnetic properties of PtSn$_{4}$ are consistent with it being a weakly
diamagnetic, intermetallic compound, but with a clear anisotropy. The high quality of the samples has allowed us to experimentally observe the quantum oscillations from the
magnetization (de Haas-van Alphen effect, dHvA), resistivity (Shubnikov-de Haas effect, SdH), and thermoelectric power (TEP) measurements. For the magnetic field applied along three
different orientations, we found several frequencies for each direction, implying several extremal orbits on the Fermi surface. A large magnetic field dependence of the resistivity is
observed at low temperatures in the highest quality crystals of PtSn$_{4}$, which appears to follow a generalized Kohler's rule with a functional $\Delta\rho/\rho(0) =
F[H/\rho(0)]\simeq H^{1.8}$ \cite{Pippard}. The observed transport properties derived from the Hall coefficient and TEP measurements also show a significant magnetic field dependence.

\section{Experimental Methods and Structure Determination}

Single crystals of PtSn$_{4}$ were grown out of a Sn-rich binary melt \cite{Canfield}. The constituent elements, with an initial stoichiometry of Pt$_{0.04}$Sn$_{0.96}$, were placed
in an alumina crucible and sealed in a quartz tube under a partial Ar pressure. After the quartz ampoule was heated up to 600\,$^{\texttt{o}}$C, the ampoule was cooled down to
350\,$^{\texttt{o}}$C over 60 hours. The crystals grow as large, layered-plates shown in Fig. \ref{structure} (b) and are malleable but also easily cleaved. The crystals are large,
typically up to $\sim$\,7$\times$7$\times$2 mm$^{3}$, and an almost 100\,\% yield of PtSn$_{4}$, based on initial Pt content, can be achieved. The Sn flux was removed using a
centrifuge (decanting), and the remaining Sn flux droplets on the surface were removed by etching in concentrated hydrochloric acid (HCl).

\begin{figure}
\centering
\includegraphics[width=1\linewidth]{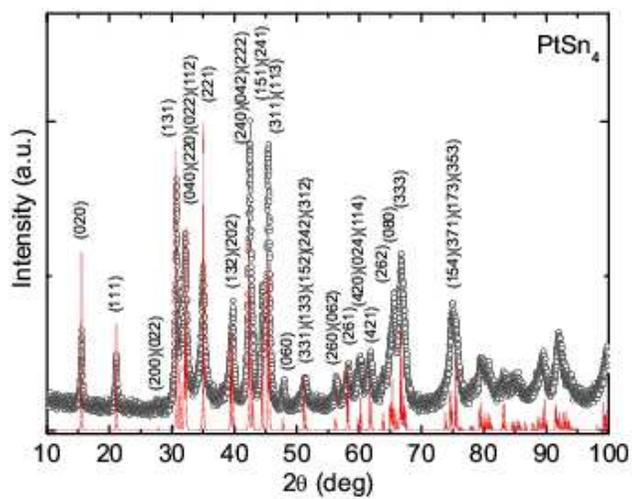}
\caption{Observed and calculated x-ray powder patterns of PtSn$_{4}$. The open circles represent the observed data points and the solid line
represent the calculated pattern. Representative ($hkl$) values are indicated.}
\label{xrd}%
\end{figure}

Powder x-ray diffraction measurements were taken at room temperature with Cu-$K_{\alpha1}$ radiation in a Rigaku Miniflex diffractometer. Because of their soft nature, the crystals
were ground for x-ray measurements in liquid nitrogen. A typical diffraction pattern of PtSn$_{4}$ is shown in Fig. \ref{xrd}. The problem of indexing and determination of the lattice
constant in powder diffraction analysis is non-trivial due fundamentally to the significant broadening of peaks, therefore lower angle peaks, below 2$\theta$ = 60$^{\texttt{o}}$, were
selected to determine lattice constants. Most of the detected peaks can be indexed using an orthorhombic (Ccca, No.68) structure with $a$ = 6.42\,$\pm$\,0.03\,\AA, $b$ =
11.42\,$\pm$\,0.05\,\AA, and $c$ = 6.39\,$\pm$\,0.03\,\AA, which gives a reasonable agreement with single crystal x-ray diffraction measurements and is consistent with the earlier
report \cite{Kunnen}. The observed and calculated patterns are shown in Fig. \ref{xrd}, where we expect that the differences between calculated and observed peak intensities are based
on preferred orientation of the powder.

In order to check for atomic positions and site occupancies in PtSn$_{4}$ and to confirm space group, multiple single crystals of approximately 0.1 $\times$ 0.1 $\times$ 0.1 mm$^{3}$
were extracted from several batches for single crystal x-ray diffraction determination of the structure. Room temperature x-ray diffraction data were collected on a STOE-IPDS2-II
(Image Plate Diffraction System) single crystal diffractometer with Mo-$K_{\alpha}$ radiation ($\lambda$ = 0.71073 \AA; 50 kV and 40 mA) using a PG(002)-monochromator. The intensity
data sets were used to solve and refine the crystal structures with the SHELXTL program suite. The final analysis of the single crystal x-ray diffraction data confirms the space group,
lattice parameters and 100\,\% site occupancy of the Pt and Sn sites, which were consistent with the earlier study \cite{Kunnen}.

The orientation of the crystal was determined by standard Laue techniques, where the \textbf{b}-axis is perpendicular to the plate and \textbf{a}-axis and \textbf{c}-axis are parallel
to the plane as well as the edge facets of the crystals. Unfortunately, due to the small difference between the $a$ and $c$ lattice constant, and somewhat poor quality of Laue
back-scattering data, it was difficult to identify specific edges as \textbf{a} or \textbf{c}. In this article we will use the orientations \textbf{b}, $\perp$\,\textbf{b}I, and
$\perp$\,\textbf{b}II where $\perp$\,\textbf{b}I and $\perp$\,\textbf{b}II are perpendicular, planar directions along the edge facets and are either the \textbf{a}-axis or
\textbf{c}-axis.

The specific heat was measured by the relaxation technique over the temperature range from 1.8 to 200\,K in a Quantum Design (QD) Physical Property Measurement System (PPMS). The
magnetization as a function of temperature and field was measured by using a QD Superconducting Quantum Interference Device (SQUID) magnetometer over the temperature range from 1.8 to
300\,K and in fields up to 70\,kOe. The electrical resistivity was measured using a four-probe, $ac$ ($f$ = 16\,Hz) technique over temperatures from 1.8 to 305\,K and in fields up to
140\,kOe in a QD PPMS. For the transverse magnetoresistance (MR) measurements, the magnetic field was applied perpendicular to the current direction, \textbf{H}\,$\perp$\,\textbf{I},
the current was flowing along the \textbf{ac}-plane of the crystal (\textbf{I}\,$\perp$\,\textbf{b}) and the magnetic field was applied along either the \textbf{b}-axis
(\textbf{H}\,$\parallel$\,\textbf{b}) or the \textbf{ac}-plane (\textbf{H}\,$\perp$\,\textbf{b}). Hall resistivity measurements were performed in a four wire geometry, switching the
polarity of the magnetic field to remove MR effects due to small misalignments of the voltage wires; the configurations of wires and fields were \textbf{H}\,$\parallel$\,\textbf{b}
and \textbf{I}\,$\perp$\,\textbf{V}\,$\perp$\,\textbf{b}. Thermoelectric power measurements were carried out by dc, alternating heating (two-heater-two-thermometer), technique
\cite{TEPsetup} over the temperature range from 2 to 300\,K and in fields up to 140\,kOe.

\section{Results}
\subsection{Specific Heat}

\begin{figure}
\centering
\includegraphics[width=1\linewidth]{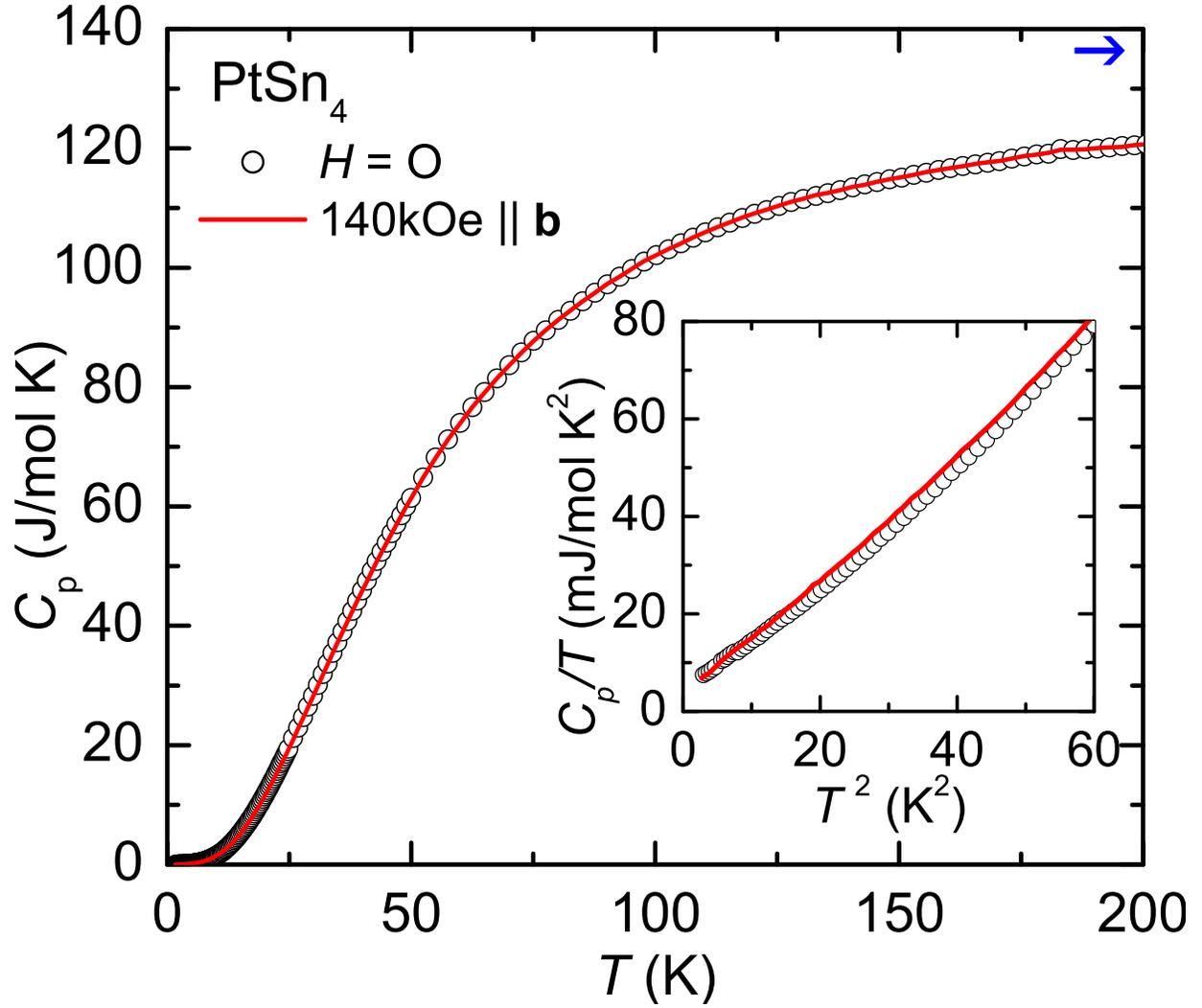}
\caption{Temperature dependence of the specific heat $C_{P}$ of PtSn$_{4}$ in $H$ = 0 (open circle) and 140\,kOe (solid line). Inset: $C_{p}/T$ vs. $T^{2}$ plot. Horizontal arrow in
the upper right corner indicates the classical limit of the Dulong-Petit law.}
\label{Cp}%
\end{figure}

The temperature-dependent specific heat, $C_{p}$, data for PtSn$_{4}$ in $H$ = 0 and 140\,kOe are presented in Fig. \ref{Cp}. The $C_{p}$ curves are typical of metallic compounds and
do not show any signature of a phase transition. The sample is essentially non-magnetic and a magnetic field of 140 kOe has essentially no effect on $C_{p}(T)$. The electronic specific
heat coefficient $\gamma$ and the Debye temperature $\Theta_{D}$ of PtSn$_{4}$ can be obtained by fitting the $C_{p}$ data to the relation $C_{p} = \gamma T + \beta T^{3}$ at low
temperatures. In zero-field, $\Theta_{D}$ is estimated to be $\sim$\,210\,K and $\gamma$ is found to be $\sim$\,4\,mJ/mol$\cdot$K$^{2}$. The specific heat increases rapidly between 25
and 75\,K and on further warming appears to be saturating toward the classical limit of Dulong-Petit law (136.6 J/mol$\cdot$K for PtSn$_{4}$) \cite{Pobell}.

\subsection{Magnetization}

\begin{figure}
\centering
\includegraphics[width=0.5\linewidth]{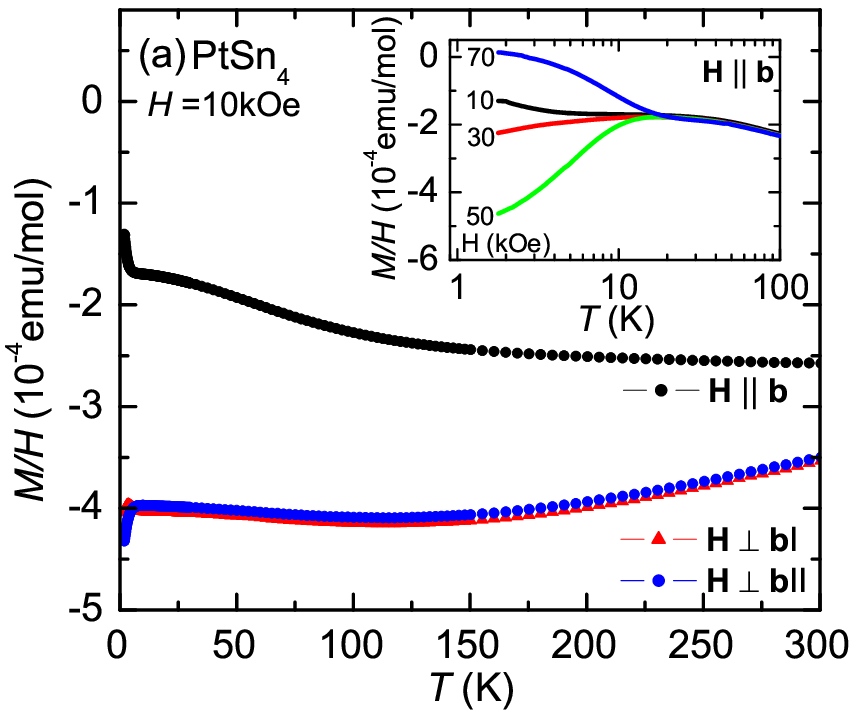}\includegraphics[width=0.5\linewidth]{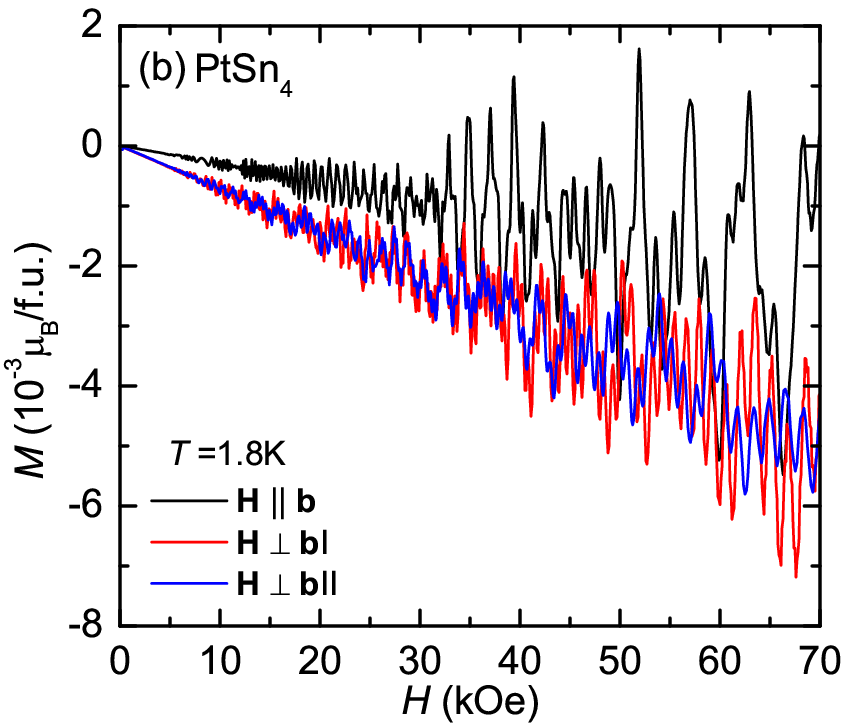}
\caption{(a) Temperature-dependent $M(T)/H$ of PtSn$_{4}$ for \textbf{H}\,$\parallel$\,\textbf{b}, \textbf{H}\,$\perp$\,\textbf{b}I, and
\textbf{H}\,$\perp$\,\textbf{b}II. Inset: Low-temperature $M(T)/H$ for \textbf{H}\,$\parallel$\,\textbf{b}. (b) Magnetization isotherms of
PtSn$_{4}$ for \textbf{H}\,$\parallel$\,\textbf{b}, \textbf{H}\,$\perp$\,\textbf{b}I, and \textbf{H}\,$\perp$\,\textbf{b}II at 1.8\,K.}
\label{MTMH}%
\end{figure}

The observed magnetic properties of PtSn$_{4}$ are consistent with it being a weakly diamagnetic, intermetallic compound. As shown in Fig. \ref{MTMH} (a), the magnetization as a
function of temperature, $M(T)/H$, measured in $H$ = 10\,kOe depends only weakly on temperature. The magnetic anisotropy is clearly seen in $M(T)/H$ between
\textbf{H}\,$\parallel$\,\textbf{b} and \textbf{H}\,$\perp$\,\textbf{b}, with essentially no anisotropy being observed within \textbf{ac}-plane (\textbf{H}\,$\perp$\,\textbf{b}I and
\textbf{H}\,$\perp$\,\textbf{b}II). At low temperatures an oscillatory behavior in $M(H)$ occurs; dHvA oscillations are shown in Fig. \ref{MTMH} (b). This non-linear low temperature
$M(H)$ is the cause of the variable $M(T)/H$ shown as the inset to Fig. \ref{MTMH} (a).

Magnetization isotherms at 1.8\,K for three different orientations are shown in Fig. \ref{MTMH} (b), where the small anisotropy between \textbf{ac}-plane and \textbf{b}-axis is
clearly revealed. dHvA oscillations in all three directions are clearly observed in field as low as 5\,kOe at $T$ = 1.8\,K, superimposed on a linear diamagnetic background. These
oscillations were detected for $H <$ 70\,kOe for temperatures as high as 30\,K. These observations imply very low conduction electron scattering and point toward the very high quality
of the samples used in the present study. Detailed analysis of the dHvA oscillations will be presented below.

\subsection{Resistivity}

\begin{figure}
\centering
\includegraphics[width=1\linewidth]{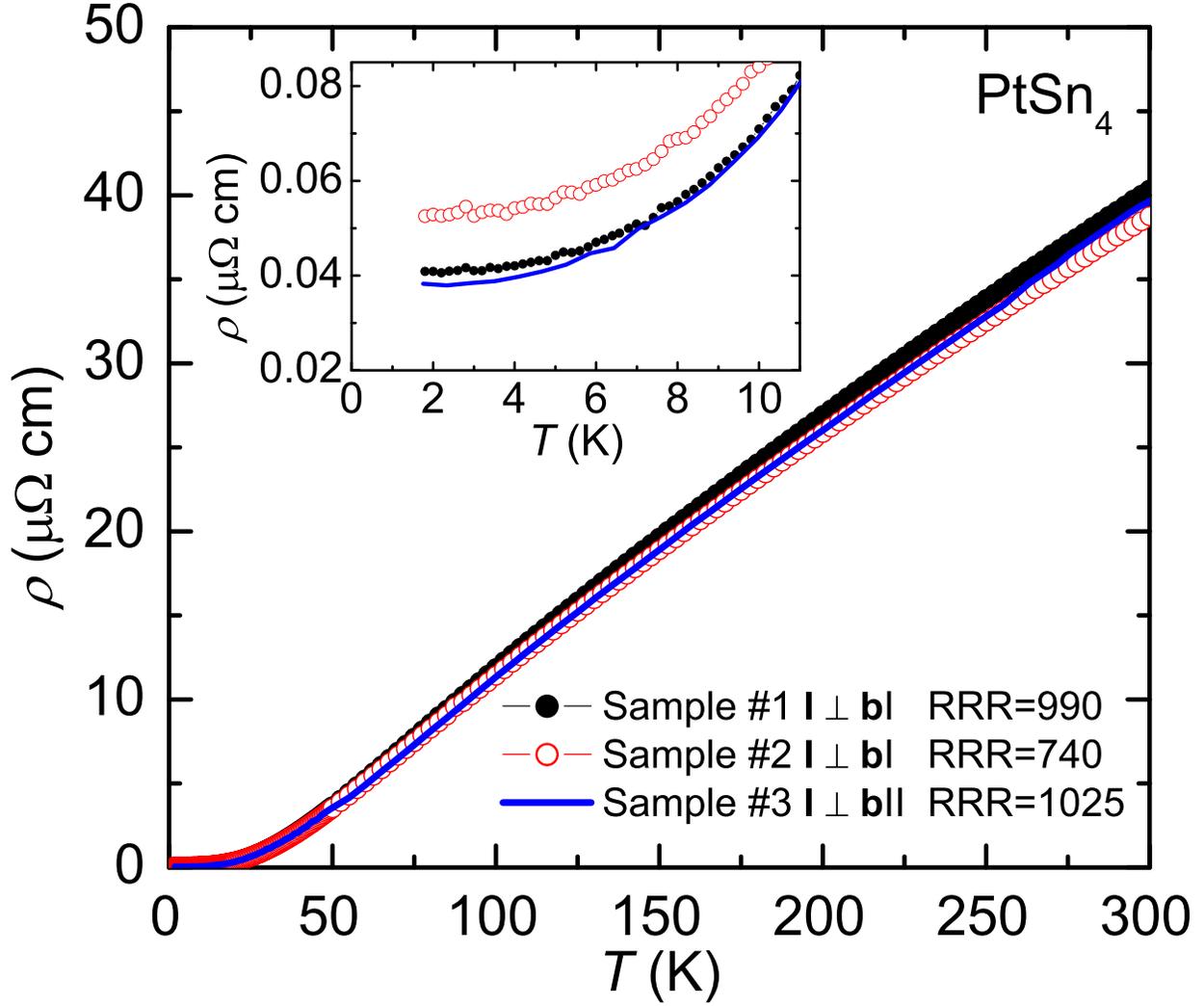}
\caption{Zero-field electrical resistivity of PtSn$_{4}$ for three samples. The current \textbf{I} was applied parallel to the plane of longer
axis (\textbf{I}\,$\perp$\,\textbf{b}I) and shorter axis (\textbf{I}\,$\perp$\,\textbf{b}II). Inset shows the low-temperature resistivity. For
details see text.} \label{RT}
\end{figure}

The electrical resistivity, $\rho(T)$, of PtSn$_{4}$ was measured from 1.8 to 305\,K and found to be metallic (Fig. \ref{RT}); for 50\,K $<$ $T$ $<$ 300\,K $\rho(T)$ decrease in a
roughly linear fashion with decreasing temperature, below 50\,K $\rho(T)$ starts to saturate with a very small residual resistivity after passing through a weak slope change, and
below 7\,K $\rho(T)$ is proportional to $T^{2}$. Figure \ref{RT} shows $\rho(T)$ of three PtSn$_{4}$ samples where the electrical current was applied parallel to the plane of
crystals, \textbf{I}\,$\perp$\,\textbf{b}, as mentioned in the experimental section. The labels \textbf{I}\,$\perp$\,\textbf{b}I and \textbf{I}\,$\perp$\,\textbf{b}II indicate that the
applied current is parallel to the longer and shorter orthogonal edge facets of the crystal, respectively. All three samples have similar values of the resistivity at 300\,K as well
as similar residual resistivities, indicating that the \textbf{ac}-plane anisotropy of PtSn$_{4}$ is small. The residual resistivity, $\rho_{0}$, and the residual resistivity ratio
(RRR) determined by $\rho$(300 K)/$\rho$(2 K) are $\rho_{0}$ = 0.041\,$\mu\Omega$cm and RRR = 740, $\rho_{0}$ = 0.053\,$\mu\Omega$cm and RRR = 990, and $\rho_{0}$ =
0.038\,$\mu\Omega$cm and RRR = 1025 for samples \#1, \#2 and \#3, respectively. The very low residual resistivity and the large RRR indicate the exceptional quality of samples. The
power law analysis of the resistivity with the relation $\rho(T) = \rho_{0} + AT^{2}$ at low temperatures results the small coefficient of $A$ $\simeq$ 2$\times10^{-4}$
$\mu\Omega$cm/K$^{2}$. The small value of the electron-electron scattering coefficient, denoted by $A$, is consistent with the small effective mass estimated from specific heat
coefficient $\gamma \sim$ 4\,mJ/mol$\cdot$K$^{2}$; $A/\gamma^{2}$ $\simeq$ 1.25 $\times$ 10$^{-5}$ $\mu\Omega$ cm (mol K/mJ)$^{2}$.

\begin{figure*}
\centering
\includegraphics[width=0.5\linewidth]{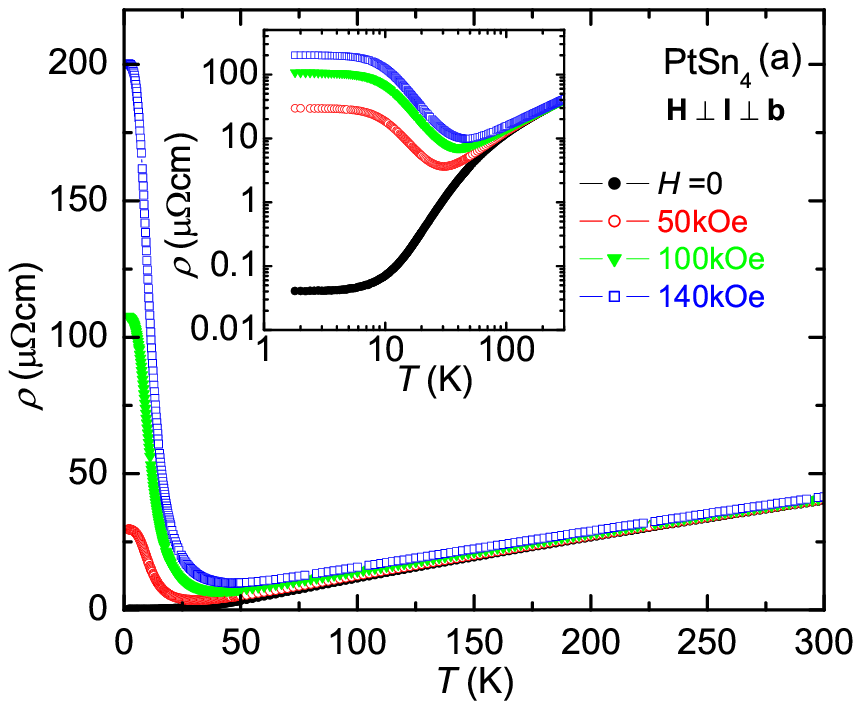}\includegraphics[width=0.5\linewidth]{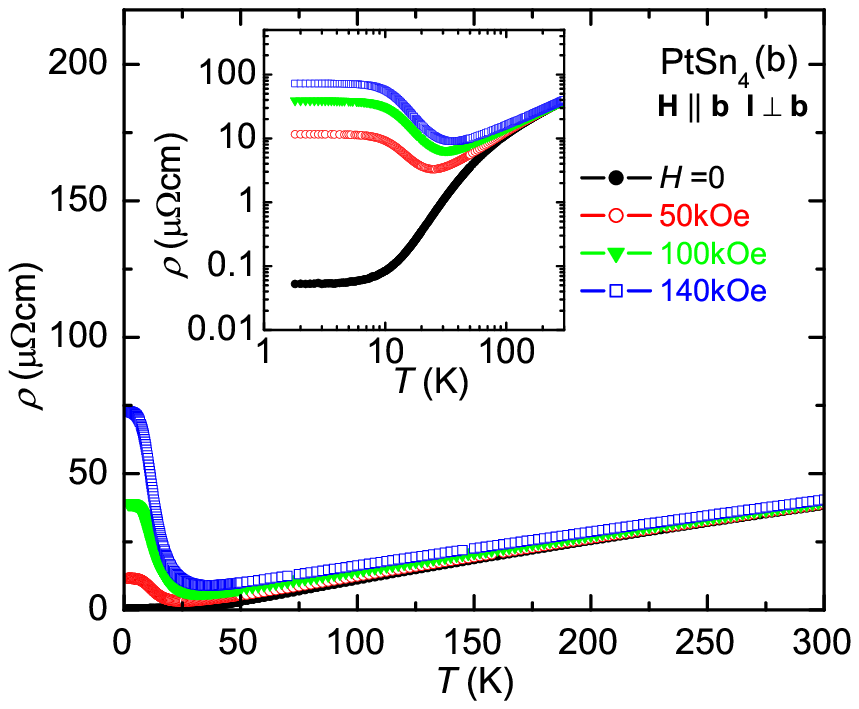}
\includegraphics[width=0.5\linewidth]{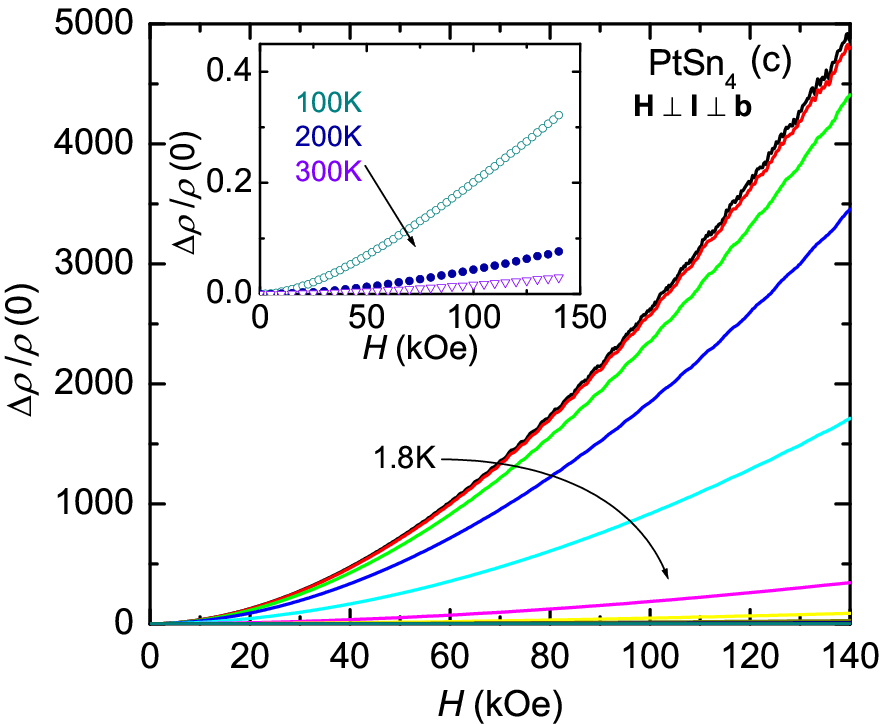}\includegraphics[width=0.5\linewidth]{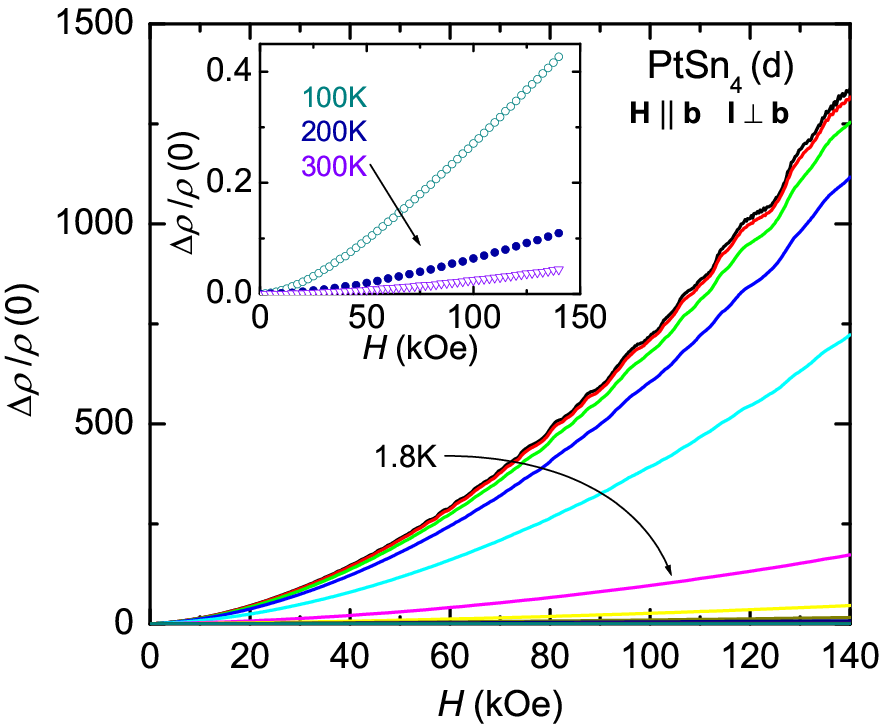}
\caption{Resistivity measurements as function of temperature and field on PtSn$_{4}$ single crystals. Temperature-dependent resistivity for (a)
an applied field along the \textbf{ac}-plane (\textbf{H}\,$\perp$\,\textbf{b}) and (b) along the \textbf{b}-axis
(\textbf{H}\,$\parallel$\,\textbf{b}). Insets (a) and (b): $\rho(T)$ vs. log($T$) plot. Magnetic field-dependent resistivity $\Delta\rho/\rho(0)$
for (c) \textbf{H}\,$\perp$\,\textbf{b} and (d) \textbf{H}\,$\parallel$\,\textbf{b} at $T$ = 1.8, 3, 5, 7, 10, 15, 20, 25, 30, 50, 100, 200, and
300\,K (top to bottom). Insets (c) and (d): $\Delta\rho/\rho(0)$ at $T$ = 100, 200, and 300\,K. The electrical current was applied in the
\textbf{ac}-plane (\textbf{I}\,$\perp$\,\textbf{b}) for all measurements.}
\label{RTRH}%
\end{figure*}

Samples were selected for transverse MR measurements with the field oriented parallel to the \textbf{b}-axis (sample \#1, \textbf{H}\,$\parallel$\,\textbf{b}) and perpendicular to the
\textbf{b}-axis (sample \#2, \textbf{H}\,$\perp$\,\textbf{b}). Figures \ref{RTRH} (a) and (b) show the temperature dependence of the resistivity at $H$ = 0, 50, 100, and 140\,kOe. The
field dependence of the resistivity for \textbf{H}\,$\perp$\,\textbf{b} is larger than that for \textbf{H}\,$\parallel$\,\textbf{b} at low temperatures: $\rho$(1.8 K, 140 kOe) for
\textbf{H}\,$\perp$\,\textbf{b} is approximately a factor of 3 larger than that for \textbf{H}\,$\parallel$\,\textbf{b}. In a given field, $\rho$($T$) increase rapidly as temperature
decreasesd from 50 to 10\,K, and then below 10\,K $\rho(T)$ is saturated to a constant value as can be more clearly seen in the log-log plot (insets). This is, of course, the same
temperature region ($T$ $<$ 10 K) that the $\rho$($T$) saturates to the residual resistivity in zero field.

As shown in Figs. \ref{RTRH} (a) and (b), the resistivity has a relatively weak field dependence above 50\,K, but depends strongly on magnetic field below this temperature. The size
and anisotropy of MR can be seen explicitly in Figs. \ref{RTRH} (c) and (d). The relative change of the MR as a function of magnetic field is plotted by the typical definition of MR
\cite{Pippard}; $\frac{\Delta\rho}{\rho(0)} = \left[\frac{\rho(H)-\rho(H=0)}{\rho(H=0)}\right]$, where $\rho (H=0)$ is the zero field resistivity for each given isotherms.
Interestingly, there is no evidence of the saturation of MR for either direction in temperature and field range measured. At $T$ = 1.8\,K and $H$ = 140\,kOe, a huge MR effect of
$\simeq$ 5$\times$10$^{5}$\,\% for \textbf{H}\,$\perp$\,\textbf{b} and $\simeq$ 1.4$\times$10$^{5}$\,\% for \textbf{H}\,$\parallel$\,\textbf{b} can be obtained. Note that at 300\,K and
140\,kOe the observed MR is small; $\Delta\rho/\rho(0)$ $\sim$ 3\% for \textbf{H}\,$\perp$\,\textbf{b} and $\Delta\rho/\rho(0)$ $\sim$ 4\% for \textbf{H}\,$\parallel$\,\textbf{b}. The
transverse MR is almost proportional to $H^{2}$ over the field range measured. The size of MR in PtSn$_{4}$ is comparable, but smaller, than Bi and much larger than that of Cu
\cite{Pippard, Ziman}. There are SdH oscillations superimposed on the MR. The amplitude of the oscillations increases with increasing field and decreasing temperature. Detailed
analysis of the SdH data will be presented below.

\subsection{Hall Coefficient and Thermoelectric Power}

\begin{figure*}
\centering
\includegraphics[width=0.5\linewidth]{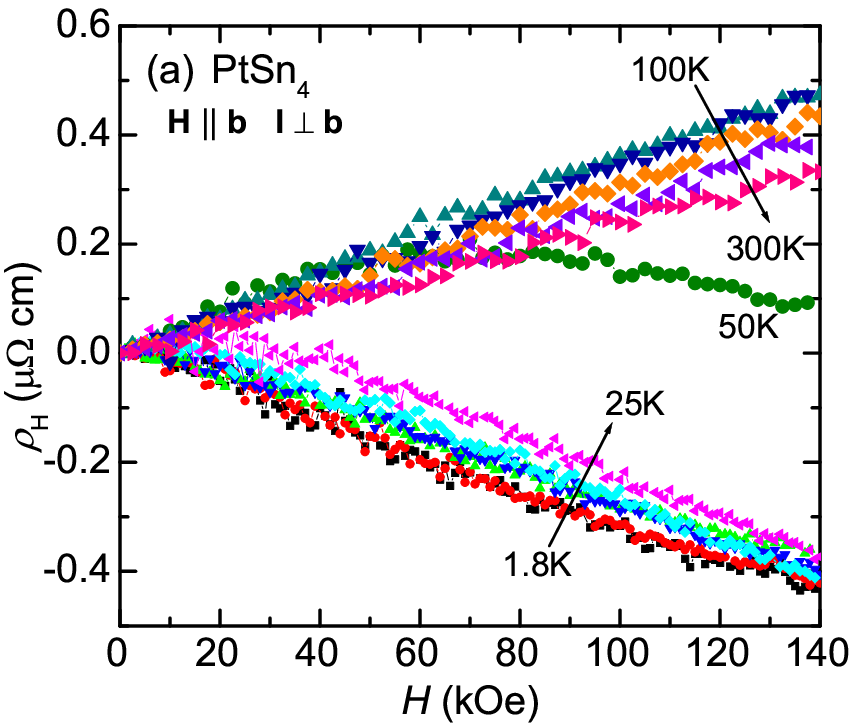}\includegraphics[width=0.5\linewidth]{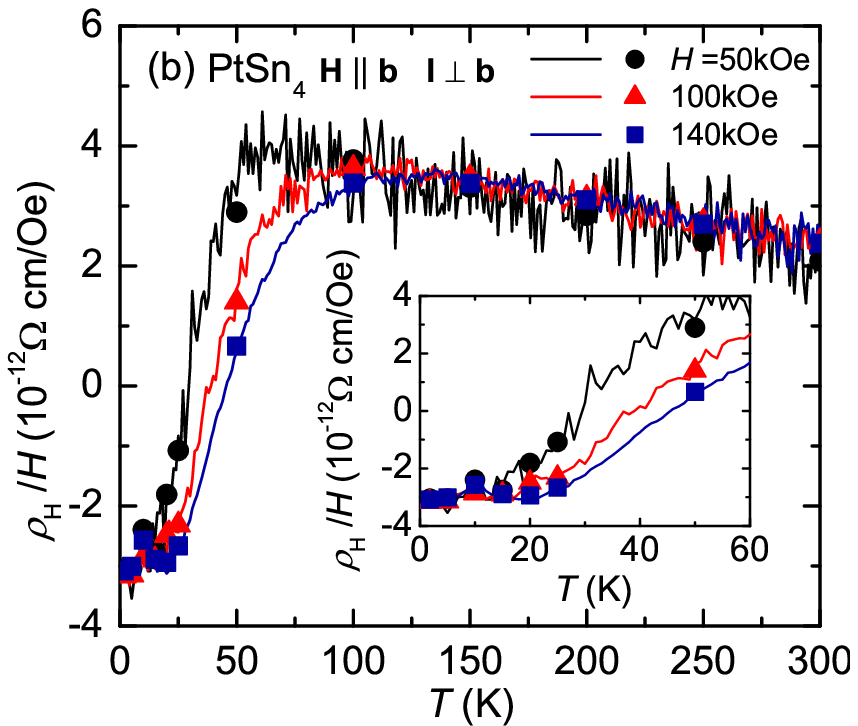}
\caption{(a) Hall resistivity ($\rho_{H}$) of PtSn$_{4}$ measured at $T$ = 100, 150, 200, 250, 300, and 50\,K (top to bottom) and $T$ = 1.8, 5,
10, 15, 20, and 25\,K (bottom to top). (b) Hall coefficient, $R_{H}=\rho_{H}/H$, of PtSn$_{4}$ at $H$ = 50, 100, and 140\,kOe. Solid lines are
obtained from the temperature dependence of $\rho_{H}$ and solid symbols are taken from the field dependence of $\rho_{H}$. Inset shows $R_{H}$
at low temperatures. The electrical current was applied in the \textbf{ac}-plane (\textbf{I}\,$\perp$\,\textbf{b}) and the magnetic field was
applied along the \textbf{b}-axis (\textbf{H}\,$\parallel$\,\textbf{b}) for all measurements.}
\label{Hall}%
\end{figure*}

The transport properties of PtSn$_{4}$ are further characterized by measurements of the Hall resistivity, $\rho_{H}$, and thermoelectric power (TEP, $S$). Figure \ref{Hall} (a) shows
the magnetic field-dependent $\rho_{H}$ measured at several temperatures. At high temperatures, $\rho_{H}$ manifests a linear field dependence with positive values. As temperature
decreases below 100\,K, through 50\,K, to 25\,K, $\rho_{H}$ becomes non-linear in magnetic field. For example, the 50\,K $\rho_{H}$ curve is positive at low fields, and then turns
towards negative values with increasing magnetic field. Below 25\,K, $\rho_{H}$ is negative and again has a linear field dependence that is comparable to the 300\,K curve.

The temperature-dependent Hall coefficient, $R_{H} = \rho_{H}/H$, of PtSn$_{4}$ is presented in Fig. \ref{Hall} (b). Above 75\,K $R_{H}$ depends weakly on temperature with a positive
value that indicates hole-like carriers dominate its electrical transport. Within the framework of an one-band approximation, the carrier density $n\simeq$ 2.6$\times$10$^{28}$
m$^{-3}$ ($R_{H}\sim$ 2.4$\times$10$^{-12}$ $\Omega$cm/Oe) at 300\,K is very close to the carrier concentration of copper $\simeq$ 8.9$\times10^{28}$ m$^{-3}$ \cite{Kittel}. Since the
measured $R_{H}$ is an average of the holes and electrons weighted by their mobilities, it will be shown below (in the discussion section), these data are treated more adequately
within the framework of a two-band model. Upon cooling below 50\,K, a sign reversal of $R_{H}$, from positive to negative, is found around 25\,K for $H$ = 50\,kOe and 45\,K for $H$ =
140\,kOe. With further cooling $R_{H}$ becomes roughly constant, saturating near $\sim$ -3$\times$10$^{-12}$\,$\Omega$cm/Oe.

\begin{figure*}
\centering
\includegraphics[width=0.5\linewidth]{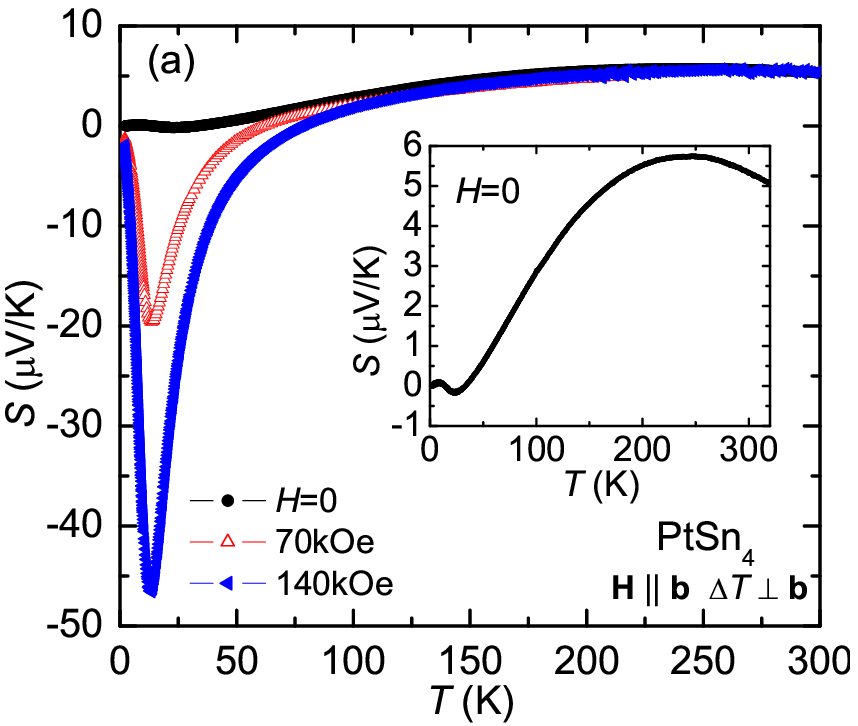}\includegraphics[width=0.5\linewidth]{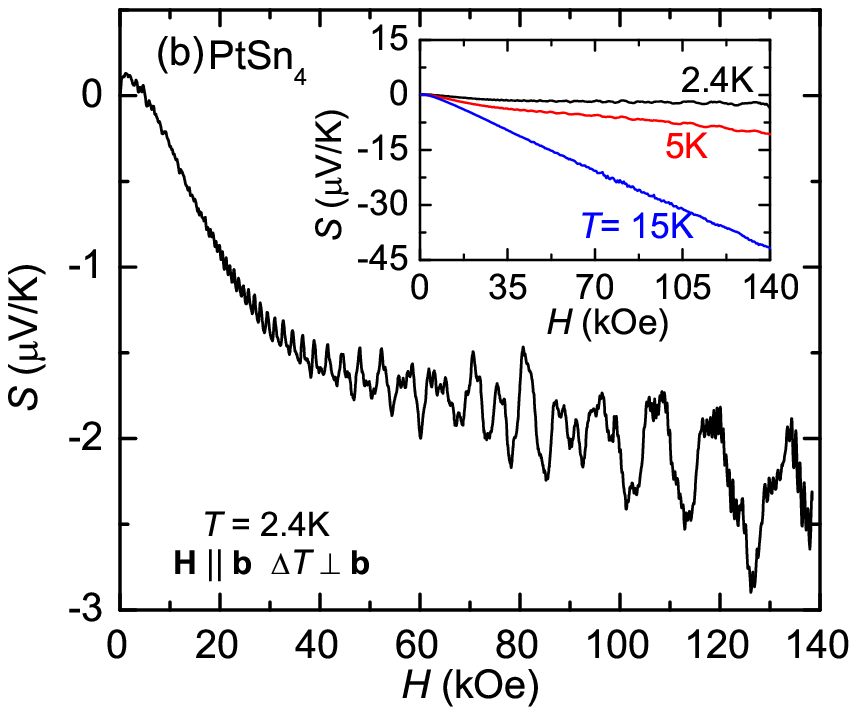}
\caption{(a) Temperature-dependent thermoelectric power (TEP, $S$) of PtSn$_{4}$ at $H$ = 0, 70, and 140\,kOe. Inset shows zero-field TEP. (b)
Magnetic field dependence of the TEP of PtSn$_{4}$ at $T$ = 2.4\,K, showing quantum oscillation for $H >$ 20\,kOe. Inset shows the TEP at $T$ =
2.4, 5, and 15\,K. The magnetic field was applied along the \textbf{b}-axis (\textbf{H}\,$\parallel$\,\textbf{b}) and the temperature difference
across the sample was generated in the \textbf{ac}-plane ($\Delta T$\,$\perp$\,\textbf{b}) for all measurements.}
\label{STSH}%
\end{figure*}

TEP data of PtSn$_{4}$ are consistent, in a qualitative way, with the behavior observed in $R_{H}$. Figures \ref{STSH} (a) and (b) show the temperature and field dependence of TEP;
$S(T, H)$. In zero field, the value of TEP is positive at high temperatures signifying that hole-type carriers dominate the thermoelectric transport in this material. Upon cooling, TEP
shows a sign reversal around 30\,K, and then exhibits a broad extremum around 20\,K ($\approx \Theta_{D}$/11) which is possibly due to the phonon-drag \cite{Blatt}. Below 50\,K,
$S(T)$ follows the same qualitative behavior as $R_{H}$, it decreases sharply and changes sign. Below 10\,K the absolute value of TEP starts to approach zero. The
magneto-thermoelectric power (MTEP) as a function of temperature is shown in Fig. \ref{STSH} (a). Above 100\,K, MTEP is positive and depends weakly on field, whereas increasingly
large MTEP effect appears below this temperature. As magnetic field increases a greatly enhanced MTEP is seen between 2 and 100\,K, dominated by the peak-like structure centered
around $\sim$ 13\,K and shifting the sign reversal temperature to higher values as field increases; changing from 33\,K for $H$ = 0 to 78\,K for $H$ = 140\,kOe. The peak position
remains the same as the magnetic field increases from 70\,kOe to 140\,kOe, located near the temperature of phonon-drag feature observed in zero field.

TEP as a function of field is plotted in Fig. \ref{STSH} (b). Quantum oscillations in the $T$ = 2.4\,K data are clearly seen for applied fields greater than 20\,kOe. At higher
temperature (inset) the absolute value of TEP increases linearly with increasing field. The quantum oscillations have been observed in the TEP of many metallic elements such as zinc
\cite{Fletcher}, semi-metal Bi \cite{Mangez}, however, compared to dHvA and SdH, have been rarely measured for intermetallic compounds \cite{Mun}.

\section{Discussion}

PtSn$_{4}$ is an ordered intermetallic compound which manifest itself as exceptionally low $\rho_{0}$, high RRR, large MR effect, and quantum oscillations found in $\rho(H)$, $M(H)$,
and $S(H)$ over wide ranges of temperature and magnetic field. The resistivity, Hall coefficient, and TEP of PtSn$_{4}$ depend strongly on magnetic field and temperature. $R_{H}$ and
$S(T)$ indicate a sign reversal with decreasing temperature, which suggests that two different types of carriers with opposite sign are present in this compound. A common
interpretation of such a sign reversal is that, as the temperature is lowered, electron and hole scattering rates change in a different manner, leading to a variation of the mobility
of charge carriers. Such a sign change of $R_{H}$ is also observed in many materials, such as Al and Bi \cite{Pippard}.

Based on the experimental results a two-band model is used to determine the main physical quantities. If there are two bands crossing the Fermi surface, the MR, Hall coefficient, and
TEP can be expressed by following equations \cite{Ziman, Sondheimer};
\begin{eqnarray}{\label{twobandmodel}} \nonumber%
\frac{\Delta\rho(H)}{\rho(0)}&=&\frac{1}{e}\frac{(1+\mu_{e}^{2}H^{2})(1+\mu_{h}^{2}H^{2})}{n_{e}\mu_{e}(1+\mu_{h}^{2}H^{2})+n_{h}\mu_{h}(1+\mu_{e}^{2}H^{2})}\\ \nonumber%
\frac{\rho_{H}}{H}&=&\frac{1}{e} \frac{n_{h}\mu_{h}^{2}-n_{e}\mu_{e}^{2}}{(n_{h}\mu_{h}+n_{e}\mu_{e})^{2}}=\frac{1}{e} \frac{n_{h}-n_{e}(\mu_{e}/\mu_{h})^{2}}{(n_{h}+n_{e}(\mu_{e}/\mu_{h}))^{2}} \\ \nonumber%
S&=&\frac{\pi^{2} k_{B}^{2} T}{3e}\left(\frac{\partial \textrm{ln}d_{0}N_{e}}{\partial E}+\frac{\partial \textrm{ln}d_{0}N_{h}}{\partial E}\right)_{E=E_{F}}\nonumber%
\end{eqnarray}
where $n_{x}$ are the carrier densities, $\mu_{x}$ the mobilities, and $N_{x}$ the density of states at the Fermi surface of electrons (e) and holes (h). The diffusion constant
$d_{0}$ in TEP is connected to the conductivity for simple form $\sigma = ne^{2}\tau/m^{*} = d_{0}N$ where $\tau$ is the relaxation time and $m^{*}$ is the effective mass of carrier.

\begin{figure*}
\centering
\includegraphics[width=1\linewidth]{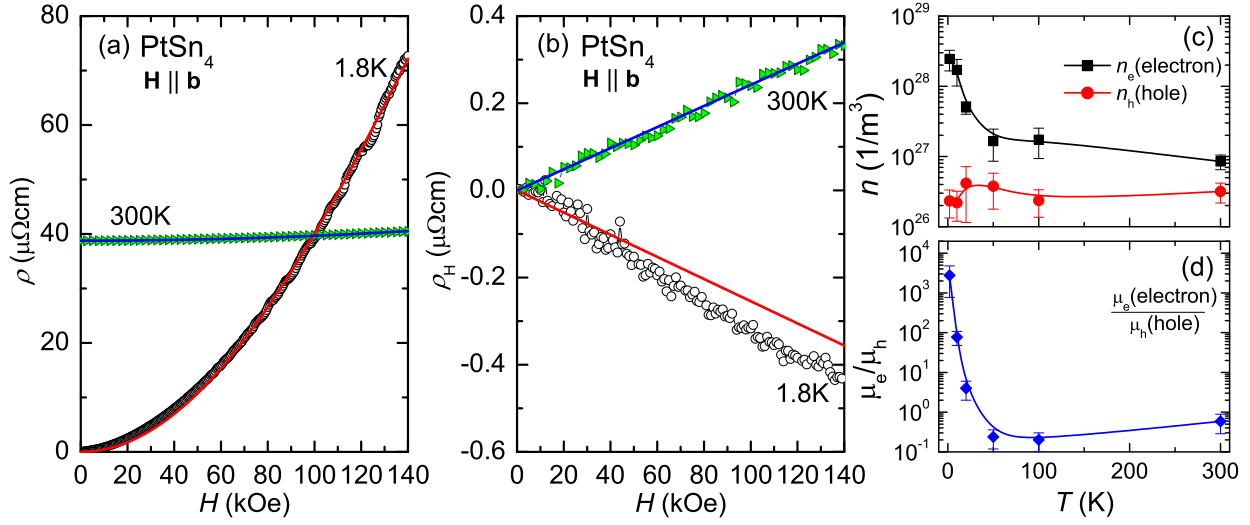}
\caption{(a) Electrical resistivity (b) Hall resistivity as a function of field at 1.8 and 300\,K for \textbf{H}\,$\parallel$\,\textbf{b}. Solid
lines are fits to the data using two-band model (see text for detailes). (c) Electron (squares) and hole (circles) concentrations as a function of
temperature, as determined from a two-band model fit to electrical resistivity and Hall resistivity. (d) Ratio ($\mu_{e}/\mu_{h}$) between the
hole and electron mobility as a function of temperature, as determined from the two-band model fit. Lines in (c) and (d) are guide to eye.}
\label{twoband}%
\end{figure*}

\begin{figure}
\centering
\includegraphics[width=1\linewidth]{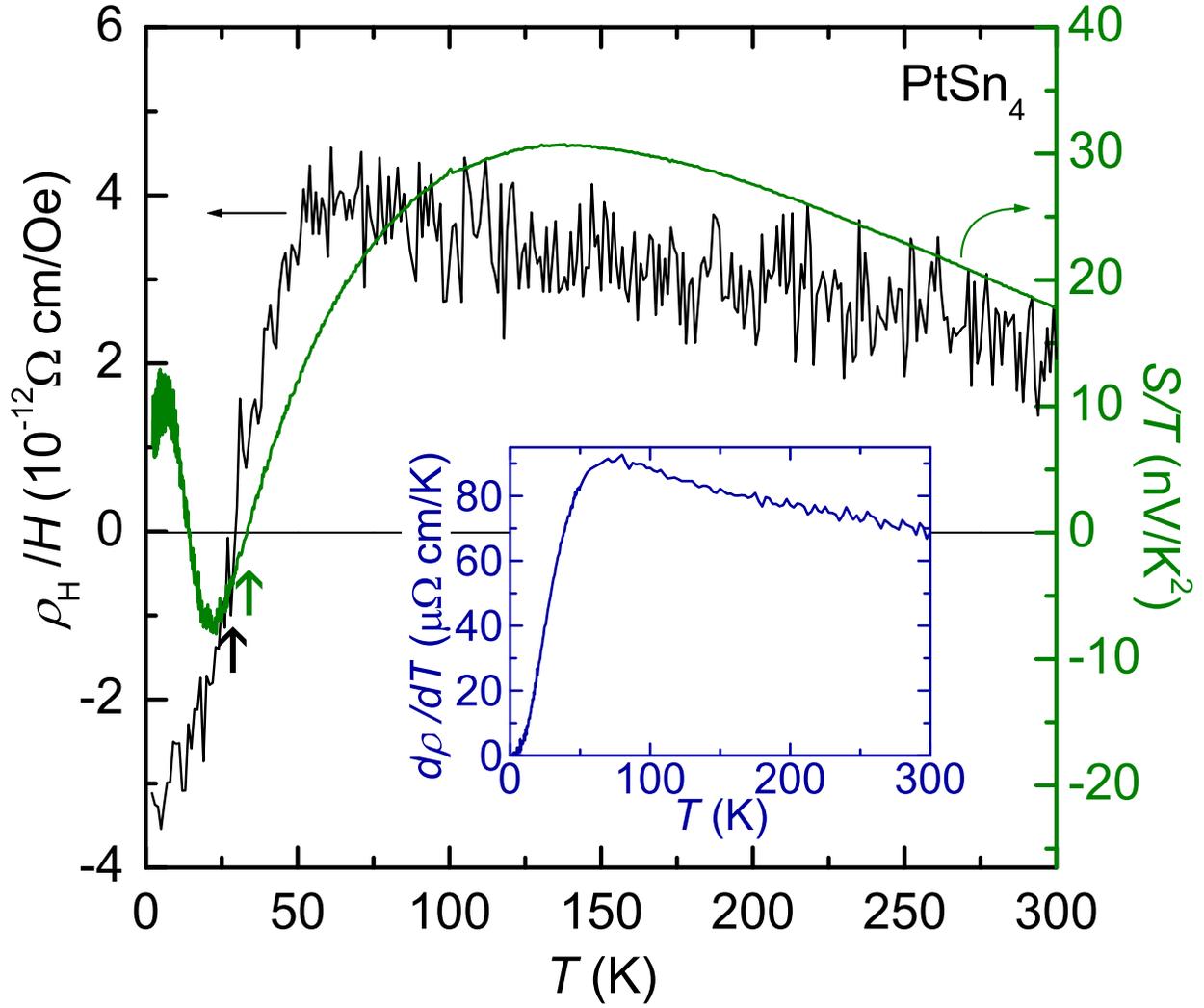}
\caption{Hall coefficient ($R_{H}$, left axis), TEP ($S(T)/T$, right axis), and resistivity ($d\rho(T)/dT$, inset) of PtSn$_{4}$ as a function of temperature. Vertical arrows indicate
the sign reversal temperature corresponding to $R_{H}$ = 0 and $S(T)/T$ = 0.}
\label{twoband1}%
\end{figure}

In the analysis of data the linear field dependence of $\rho_{H}$ and the quadratic field dependence of MR is only considered, neglecting higher order terms in magnetic field. From
the fit to the measured MR, using above equations, as shown in Fig. \ref{twoband} (a), the density of charge carriers and mobilities are estimated, and with these fit parameters
$\rho_{H}$ are reproduced as shown in Fig. \ref{twoband} (b). At 300\,K the carrier concentrations are of a similar order of magnitude, as shown in Fig. \ref{twoband} (c), but the hole
mobility is somewhat higher than the electron mobility. As the temperature is lowered below 50\,K, the electron concentration greatly exceeds the hole concentration, by two order of
magnitude, and the electron mobility is also vastly enhanced: such a temperature dependent change in carrier density could be related to the thermal depopulation of hole-like pockets
just above the Fermi level. Thus, electrons and holes are scattered in a different manner, as the temperature is lowered. As a consequence, the $\mu_{e}$/$\mu_{h}$ ratio is highly
temperature-dependent as shown in Fig. \ref{twoband} (d). This is the origin of the sign reversal in the $R_{H}$ and $S(T)$. From the above equations in the two-band model $S(T)/T$ is
expected to be similar to $R_{H}$. The temperature dependences of the $S(T)/T$ and $R_{H}$ are shown in Fig. \ref{twoband1}. This plot shows that the transport properties are
dominated by hole-like carriers at high temperature and by electron-like carriers at low temperature. The low temperature upturn in $S(T)/T$ is related to the phonon-drag. The large
change in $S(T)/T$ and $R_{H}$ involving a sign reversal around 30\,K is caused by the enhancement of the ratio $\mu_{e}/\mu_{h}$. Note that the resistivity also shows a drastic
change of scattering mechanism in the same temperature region which can be seen in $d\rho(T)/dT$ curve, as shown in the inset of Fig. \ref{twoband1}.

\begin{figure}
\centering
\includegraphics[width=1\linewidth]{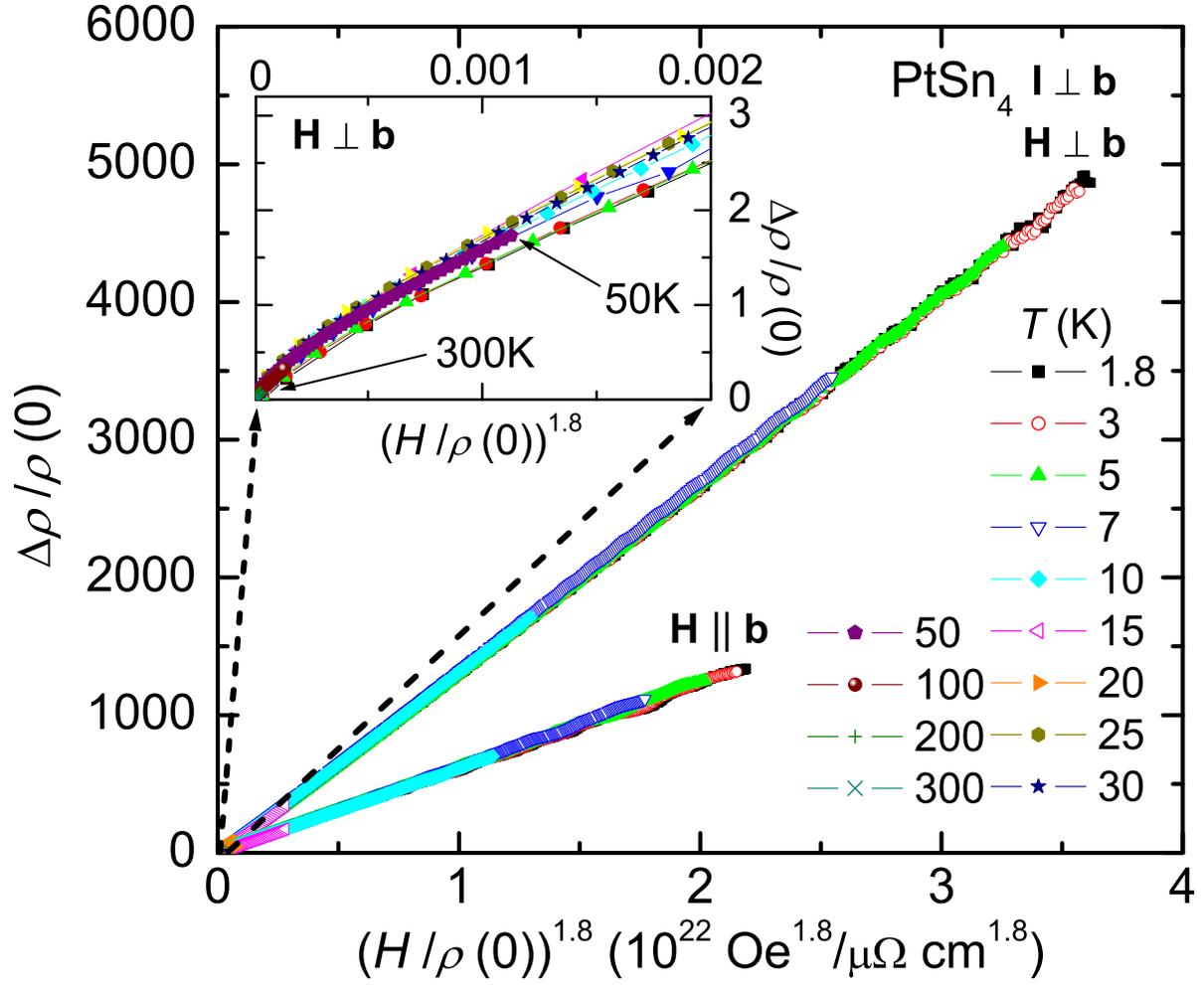}
\caption{A Kohler plot for PtSn$_{4}$ using $\Delta\rho/\rho(0) = F[H/\rho(0)]\simeq H^{1.8}$. Inset shows a Kohler plot at high temperatures for
\textbf{H}\,$\perp$\,\textbf{b}.}
\label{Kohler}%
\end{figure}

In a simple case of a metal with electrons and holes, one may find the simple relation $\Delta\rho/\rho(0) \sim (\omega_{c}\tau)^{2}$ where $\omega_{c} = eH/m^{*}c$ is the cyclotron
frequency. This relation also holds for PtSn$_{4}$. From the Fig. \ref{RTRH}, no clear indication of saturation in MR is observed either field orientation,
\textbf{H}\,$\parallel$\,\textbf{b} or \textbf{H}\,$\perp$\,\textbf{b}. The relative MR of many metals and semimetals can be represented by the form, commonly referred to as Kohler's
rule \cite{Pippard}, $\Delta\rho/\rho(0) = F[H/\rho(0)]$, where $F(H)$ usually follows a power-law, depending on the geometrical configurations and on the materials. The normal MR in
fixed orientation of the crystal either saturates at high fields or increases as $H^{2}$ \cite{Pippard, Ziman}. As shown in Fig. \ref{Kohler}, PtSn$_{4}$ has a field dependence of
$\Delta\rho/\rho(0) \simeq H^{1.8}$ for both orientations of field, where all resistivity curves collapse onto a single line for each orientation. At high temperatures and very
low-field regime at low temperatures, although MR curves show a scaling behavior, $\Delta\rho/\rho(0)$ curves do not follow $H^{1.8}$-dependence shown in the inset of Fig.
\ref{Kohler} for \textbf{H}\,$\perp$\,\textbf{b}.

In the two-band model, the experimental MR and Hall data can be explained by the $\mu_{e}$/$\mu_{h}$ ratio and carrier density, where $n_{e}$ shows a peculiar behavior;
$n_{e}\,\simeq\,n_{h}$ at 300\,K and $n_{e}\,\gg\,n_{h}$ at 2\,K. Below 30\,K $\rho_{H}/H$ data are field-independent and MR curves can be explained well with Kohler's rule. At 2\,K
the estimated value of $n_{e}$ = 2.4$\times$10$^{28}$\,m$^{-3}$ is close to the carrier concentration $n$ $\sim$ 2$\times$10$^{28}$\,m$^{-3}$ in a single-band approximation. Hence, at
the base temperature we can assume that PtSn$_{4}$ is like a single-band (electron-like) metal. Based on this assumption, the large MR effect in PtSn$_{4}$ can be understood
qualitatively by considering the quantity $\omega_{c}\tau$ as well as the mean free path ($l$). For the normal MR it is important to consider the quantity $\omega_{c}\tau =
H\sigma/nec$ which means the angle turned between collisions when a magnetic field is large enough to bend the trajectory of carriers into helices due to the Lorentz force. It has
been shown that in order to have a large MR effect, generally the criterion $\omega_{c}\tau \gg$ 1 should be satisfied \cite{Pippard}. At low temperatures it is clear from the small
residual resistivity ($\rho_{0}$ $\sim$ 0.04\,$\mu\Omega$ cm) as well as the clear and persistent dHvA oscillation that the mean free path of carriers near the Fermi surface must be
long. At the first stage the mean free path can be estimated by using the simple approximation $\rho l$ = $(3\pi^{2}/n^{2})^{1/3}\hbar/e^{2}$ \cite{Pippard} with $n$ of carriers
obtained from $R_{H}$. At 2\,K the estimated mean free path of carriers is about 3\,$\mu$m, and at 300\,K, the mean free path of carriers is much shorter than 2\,K, estimated to be
$\sim$ 3\,nm. The size of MR effect for \textbf{H}\,$\parallel$\,\textbf{b} is determined by $\omega_{c}\tau \simeq$ 6.2$\times$10$^{-4}$$H$ at 300\,K and $\omega_{c}\tau \simeq$
0.6$H$ at 2\,K; the $\omega_{c}\tau$ at 2\,K is about 10$^{3}$ time higher than $\omega_{c}\tau$ at 300\,K, the carriers are turned easily at 2\,K between collisions which is enough to
change the conduction process. Hence, the key to realizing a large MR effect of PtSn$_{4}$ is the large $\omega_{c}\tau$, which is directly related to the sample quality. Although a
single (spherical)-band approximation may not be valid for PtSn$_{4}$ system, the large MR effect can be understood in a qualitative way using single-band model, fundamentally due to
the unusually high $\mu_{e}$/$\mu_{h}$ ratio and $n_{e} \gg n_{h}$ below 30\,K. In comparison, the carrier concentration of PtSn$_{4}$ is similar to Cu, but $m^{*}$ determined by dHvA
and SdH oscillation (see below) is an order of smaller than Cu, and $\omega_{c}\tau$ is about two orders of magnitude larger than Cu \cite{Seitz}. Note that, although the temperature
dependence of transport properties in PtSn$_{4}$ can be explained qualitatively in two-band model, the temperature dependence of the carrier concentration below 50 K (in two band
model calculation) is unusual, so the realistic band structure and a three or four band fit need to be considered.

The observed behavior of $\Delta\rho/\rho(0) \sim H^{2}$ and $R_{H} \sim H^{1}$ is close to what is expected for an ordinary metallic compound with no open orbits. Although the MR is
not simply related to the topology of the Fermi surface, MR data measured for both \textbf{H}\,$\parallel$\,\textbf{b} and \textbf{H}\,$\perp$\,\textbf{b} indicate a significant
anisotropy in the band structure. In a compensated metallic compounds, no saturation is expected to be observed for either the transverse or longitudinal MR, where MR shows a
deviation from quadratic field dependence. For instance, the transverse MR of the $R$AgSb$_{2}$ ($R$ = rare-earth) series shows a deviation from quadratic field dependence, where
$\Delta\rho/\rho(0)$ ranges between $H^{0.8}$ and $H^{1.5}$ \cite{Myers}. In a similar manner the $R$Sb$_{2}$ compounds have power laws of $\Delta\rho/\rho(0)$ varying from 1 to 1.3
\cite{Bud'ko}.

The size of MR of PtSn$_{4}$ at low temperatures is comparable to or even greater than the largest giant magnetoresistance (GMR) \cite{Baibich, Berkowitz, Xiao} and colossal
magnetoresistance (CMR) \cite{Searle, von Helmol, Chahara, Jin, Tokura} materials. One of the several technological applications based on large MR effects at low temperatures will be
the magnetic field sensor. The MR effect in PtSn$_{4}$ may be suitable for detecting relatively small magnetic fields or field changes for fields up to, at least 140\,kOe. At 1.8\,K
and 140\,kOe, a huge MR effect of $\simeq$ 5$\times$10$^{5}$\,\% for \textbf{H}\,$\perp$\,\textbf{b} and $\simeq$ 1.4$\times$10$^{5}$\,\% for \textbf{H}\,$\parallel$\,\textbf{b} can
be obtained. If we assume that the observed MR continues without saturation, the very high sensitivity and a large dynamic range (below 30\,K) can potentially be used to detect
magnetic fields generated by pulsed fields ($H >$ 30\,T) and superconducting magnets (20\,T). Among intermetallic compounds, the highly anisotropic LaSb$_{2}$ \cite{Bud'ko} has been
proposed to use as a high magnetic field sensor in either a transverse MR or a Hall configuration \cite{Young}. It needs to be noted that, when intermetallic compounds are used as a
magnetic field sensor, in particular, in pulsed fields, (also oscillations are not good for sensor), eddy current heating needs to be considered.

In addition to the large MR effect, TEP also shows a large magnetic field dependence. Figure \ref{STSH} shows that, in the presence of magnetic field, a large enhancement of $S(T)$ for
PtSn$_{4}$ is observed at low temperatures. Such a large change of MTEP with applied field is reported in few materials, such as noble metals \cite{Blatt}, doped semiconductors InSb
and AgTe \cite{Heremans, Sun}, and materials with charge density wave order LaAgSb$_{2}$ \cite{Mun}. For the semiconductor InSb \cite{Heremans}, a very large MTEP was induced by the
different mobility between minority and majority carriers when acoustic phonon scattering dominates. Additionally, the GMR multilayers, granular films \cite{Shi}, and the CMR
manganites \cite{Jaime} have been also reported to have large MTEP. Among these examples, the observed absolute values of $\Delta S$ = $S(H)$ - $S(0)$ in Ag$_{2-\delta}$Te \cite{Sun},
which are an order of higher than other materials, suggest that this effect may be basically due to disorder and band-crossings. Compared with those large MTEP materials, the enhanced
TEP of PtSn$_{4}$ in magnetic field can be understood by considering the mobility change of charge carriers with electron-phonon scattering. It can be supported by the fact that $S(T)$
at 70\,kOe and 140\,kOe shows the peak-like structure near $\Theta_{D}$/11 and a similar temperature dependence between $S(T)/T$ and $R_{H}$ can be explained with two-band model (Fig.
\ref{twoband1}).

\begin{figure*}
\centering
\includegraphics[width=0.45\linewidth]{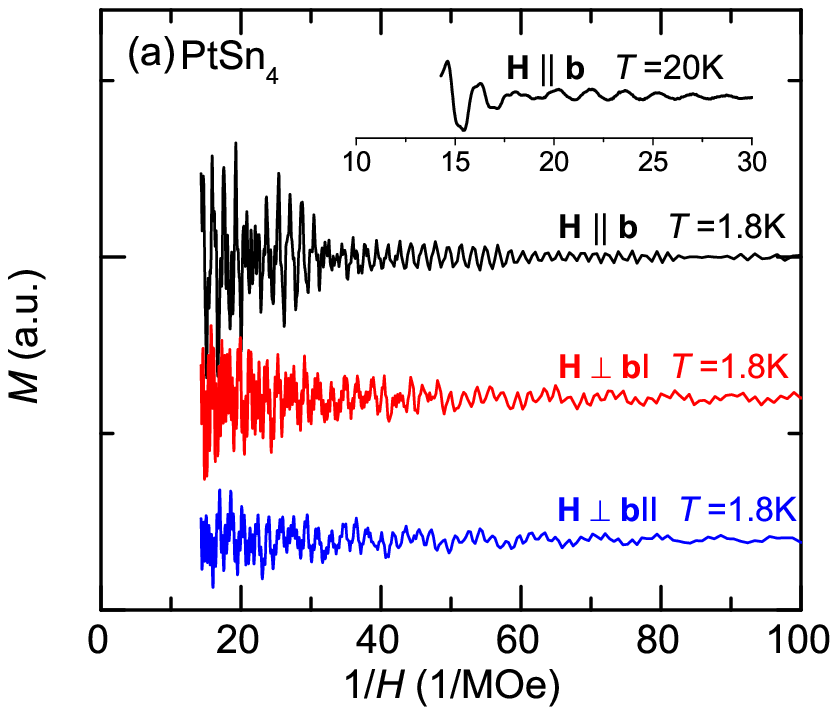}\includegraphics[width=0.45\linewidth]{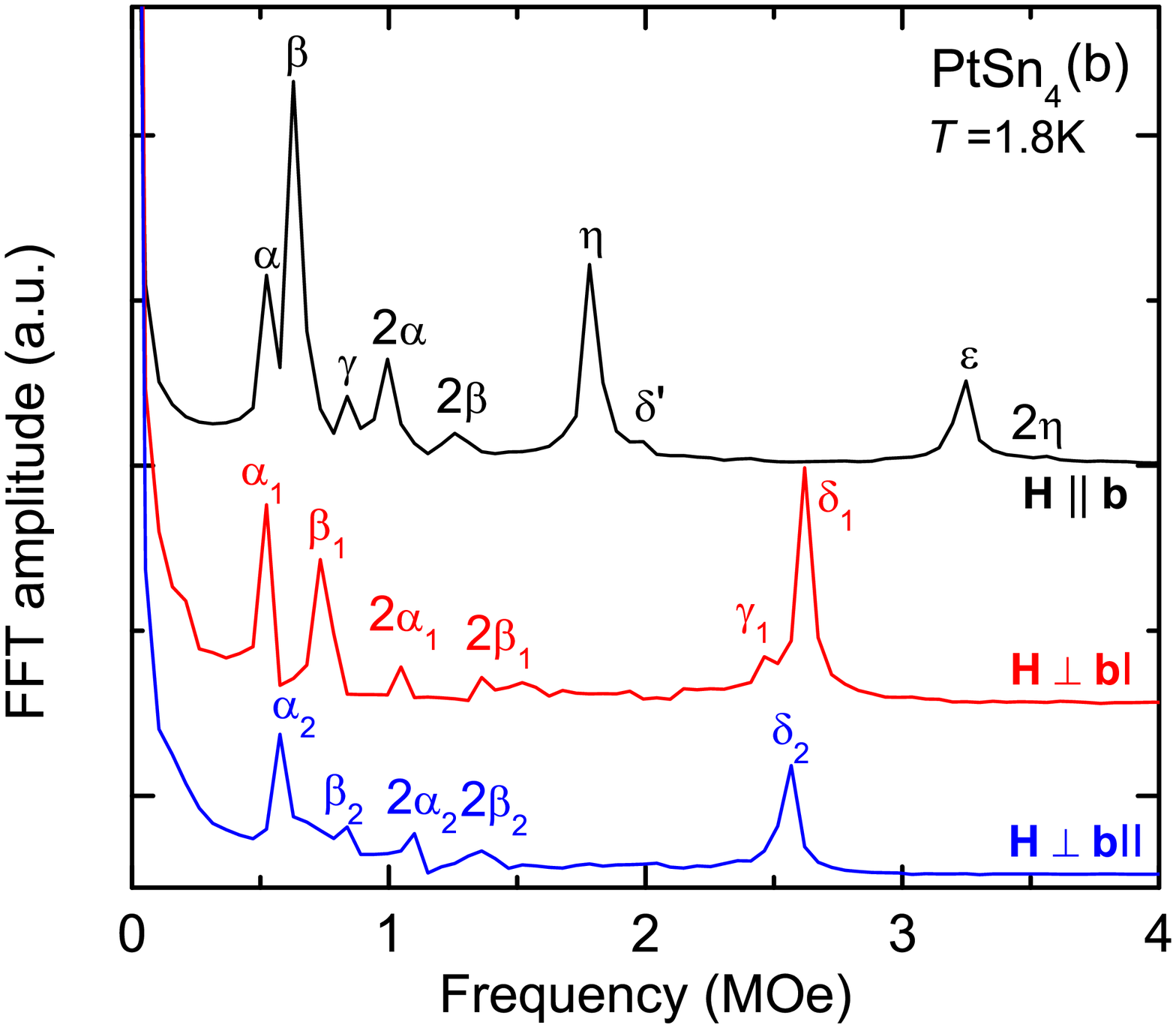}
\includegraphics[width=0.45\linewidth]{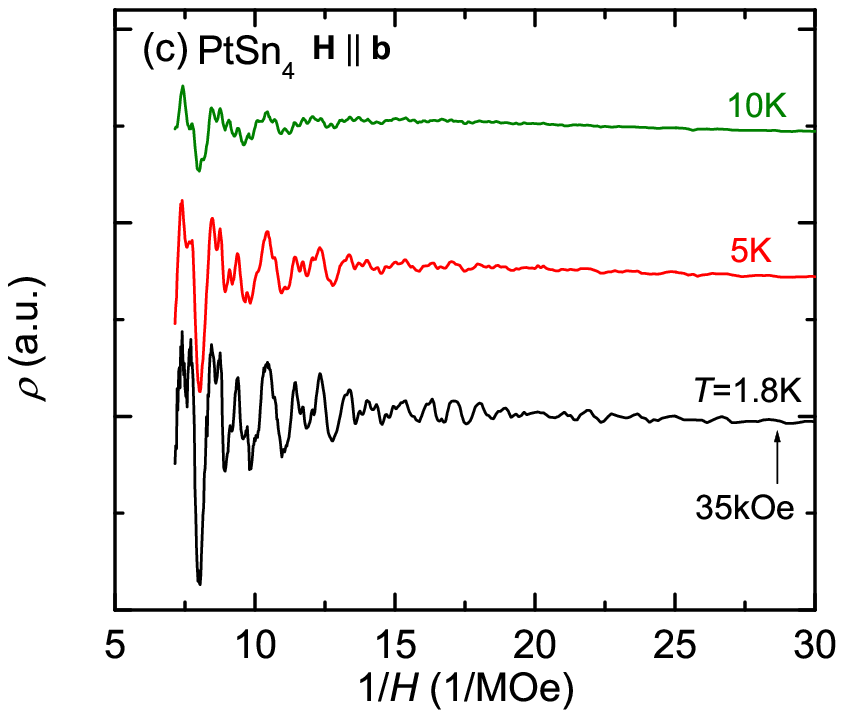}\includegraphics[width=0.45\linewidth]{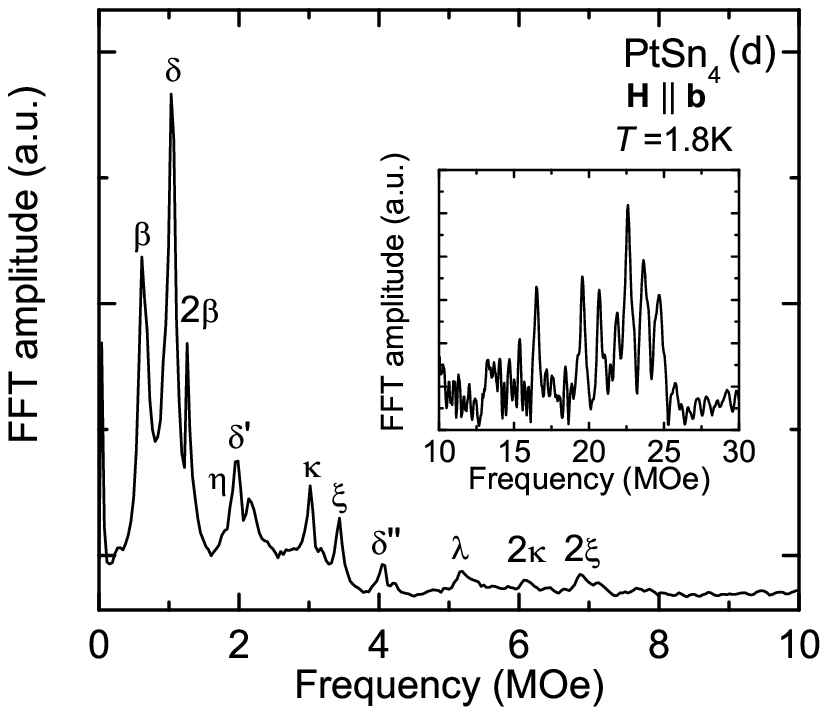}
\includegraphics[width=0.45\linewidth]{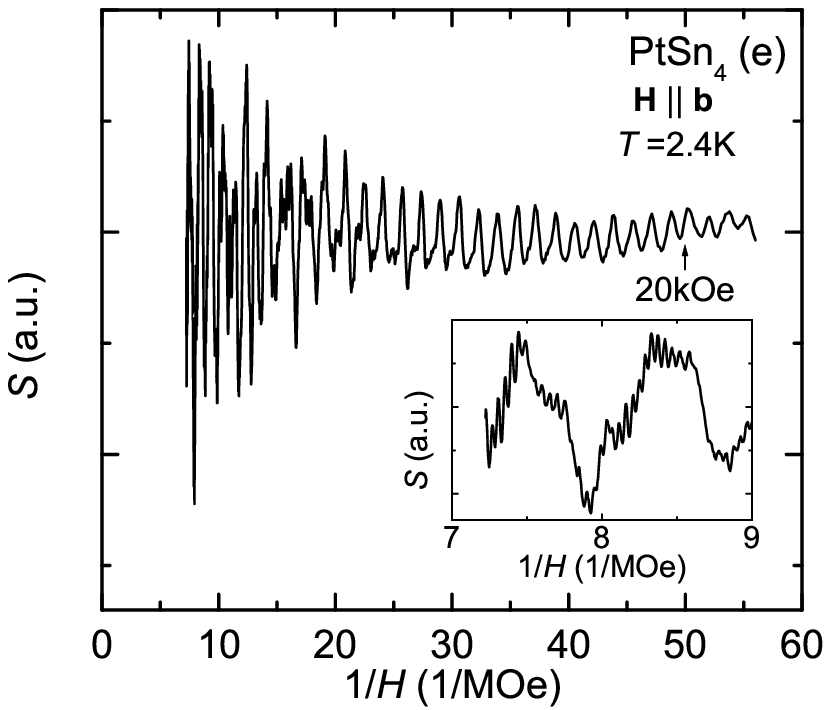}\includegraphics[width=0.45\linewidth]{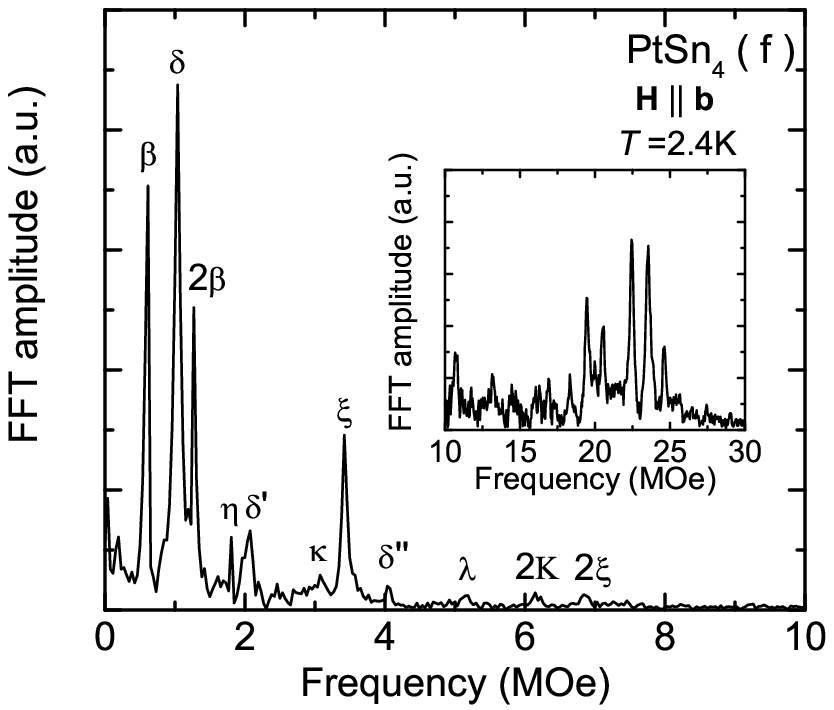}
\caption{(a) Magnetization isotherms of PtSn$_{4}$ at $T$ = 1.8\,K for three different orientations, plotted as a function of $1/H$. Inset shows a
magnetization isotherm at 20\,K for \textbf{H}\,$\parallel$\,\textbf{b}. (b) FFT spectra of dHvA data for three different orientations. (c)
Resistivity of PtSn$_{4}$ at 1.8, 5, and 10\,K for \textbf{H}\,$\parallel$\,\textbf{b}. (d) FFT spectra of SdH data at $T$ = 1.8\,K. Inset shows
the FFT spectra in the high frequency region. (e) TEP of PtSn$_{4}$ at $T$ = 2.4\,K for \textbf{H}\,$\parallel$\,\textbf{b}. Inset shows the TEP
in the high field region. (f) FFT spectra of TEP data at $T$ = 2.4\,K. Inset shows the FFT spectra in the high frequency region.}
\label{FFT}%
\end{figure*}

Oscillatory effects have been observed throughout our measurements, at remarkably low fields and high temperatures. In Fig. \ref{FFT} (a) typical dHvA data sets for PtSn$_{4}$ taken
at 1.8\,K for the three different directions of applied field are displayed, where the linear background magnetization has been subtracted. The dHvA oscillations are observed over a
wide range of temperatures for all field directions. The inset of Fig. \ref{FFT} (a) shows an example of the dHvA oscillation for \textbf{H}\,$\parallel$\,\textbf{b} at 20\,K. The fast
Fourier transform (FFT) of these data, shown in Fig. \ref{FFT}(b), reveals that these oscillations result from several independent frequencies in 1$/H$, indicating many extremal Fermi
surface cross sectional areas. The obtained frequencies are summarized in Table \ref{table1}. For \textbf{H}\,$\parallel$\,\textbf{b}, five-frequencies can be clearly observed in the
FFT spectrum corresponding to strong peaks at $\alpha$, $\beta$, $\gamma$, $\eta$, and $\varepsilon$. The second harmonics of the frequencies $\alpha$, $\beta$, $\eta$, and
$\varepsilon$ and other frequencies $\delta$' and $\delta$'' are also present as very weak peaks. FFT spectra of the oscillatory magnetization for applied fields perpendicular to the
\textbf{b}-axis at 1.8\,K are shown in the same figure. For \textbf{H}\,$\perp$\,\textbf{b}I three clear peaks and several weak signals are visible at $\alpha_{1}$, $\beta_{1}$,
$\gamma_{1}$, $\delta_{1}$, $2\alpha_{1}$, $2\beta_{1}$, $2\gamma_{1}$, and $2\delta_{1}$. The frequencies of the oscillations observed for \textbf{H}\,$\perp$\,\textbf{b}II are
similar to those observed for \textbf{H}\,$\perp$\,\textbf{b}I.

Quantum oscillations are also clearly observed in transport measurements. Figure \ref{FFT} (c) shows the resistivity data as a function of $1/H$ at $T$ = 1.8, 5, and 10\,K for
\textbf{H}\,$\parallel$\,\textbf{b}. SdH oscillations are clearly seen after subtracting the background MR ($\propto H^{1.8}$). Since the signals are comprised of a superposition of
several oscillatory components the data are most easily understood by taking the FFT as shown in Fig. \ref{FFT} (d). The FFT spectrum at $T$ = 1.8\,K shows several frequencies
including second harmonics. Additionally, several higher frequencies with smaller amplitude are resolved near 25\,MOe as shown in the inset of Fig. \ref{FFT} (d). Figure \ref{FFT} (e)
shows the TEP data as a function of $1/H$, subtracted background contributions using 3-rd order polynomial fit, taken at $T$ = 2.4\,K for \textbf{H}\,$\parallel$\,\textbf{b}. The TEP
has some distinct differences associated with quantum oscillations when compared with more conventional thermodynamic and transport measurements. Since specific heat, magnetization,
resistivity, \textit{etc}, all depend on density of states, quantum oscillations are associated with the periodic variation of this quantity. The unique aspect of TEP is that it
depends on the derivative of the density of states evaluated at the Fermi energy \cite{Fletcher, Mun}. Oscillations in TEP are clearly seen for $H >$ 20\,kOe, where higher frequencies
superimposed on lower frequencies are discernible for $H >$ 100\,kOe as seen in the inset. In Fig. \ref{FFT} (f) the FFT spectrum shows clearly three frequencies with higher amplitude
and several frequencies with lower amplitude. Besides these intense features and their harmonics, several other frequencies with much smaller amplitude are observed near 25\,MOe as
shown in the inset of Fig. \ref{FFT} (f), which is consistent with SdH results (inset, Fig. \ref{FFT} (d)). At least five frequency branches are visible, which will be related to the
very large extremal cross-sectional areas of the Fermi surface. Branches of frequency below 10\,MOe, clearly revealed in SdH and TEP oscillations, are listed in Table \ref{table1}.

\begin{figure}
\centering
\includegraphics[width=1\linewidth]{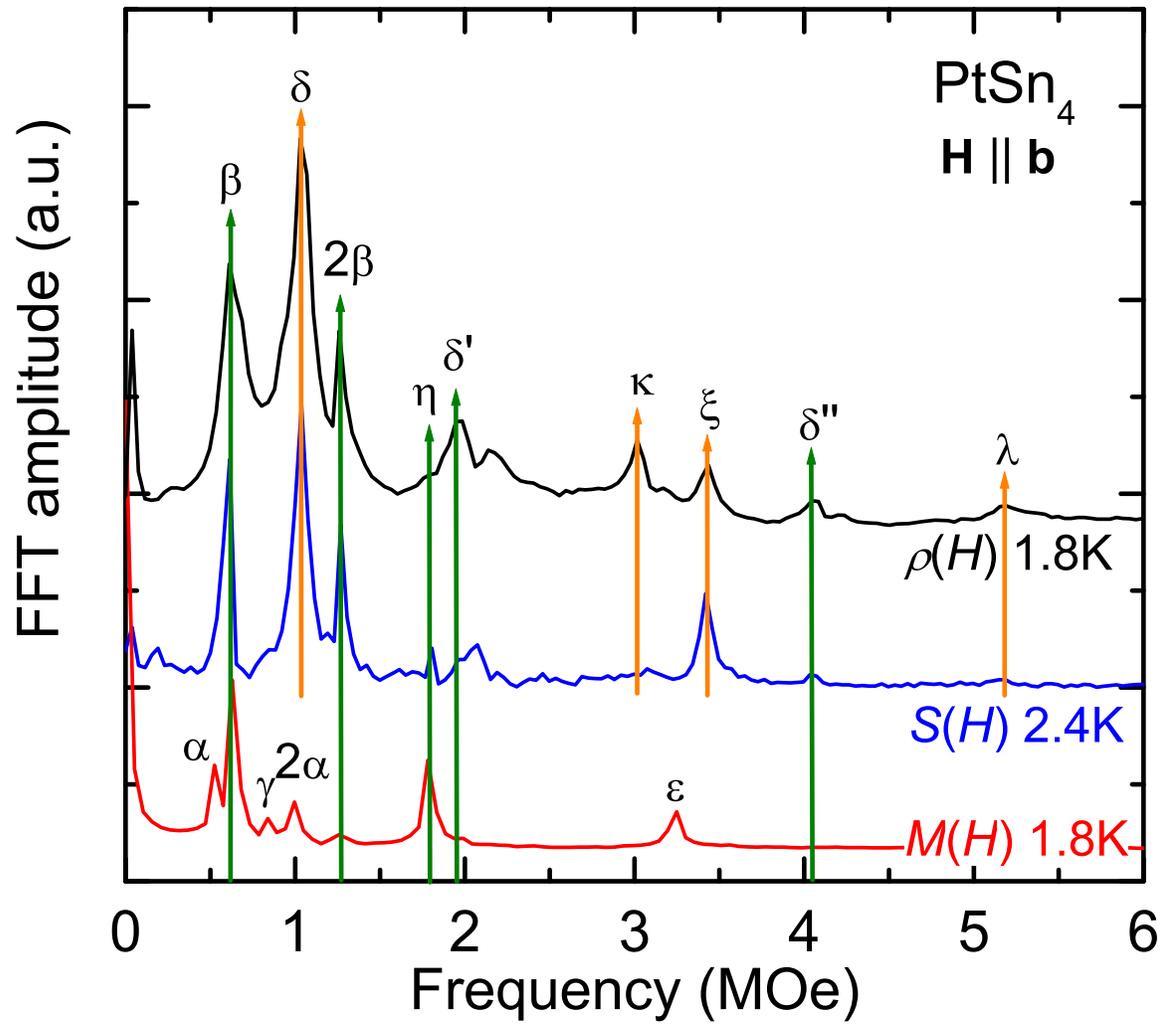}
\caption{FFT spectra of the dHvA, SdH, and TEP data. Vertical arrows are guide to eye.}
\label{FFTall}%
\end{figure}

The FFT spectra obtained from dHvA, SdH, and TEP are also plotted in Fig. \ref{FFTall} for comparison. As shown in the figure frequencies of $\beta$, $\eta$, $\delta$', and $\delta$''
in the FFT spectra corresponding to the SdH and TEP oscillation are consistent with the dHvA frequencies, where the high frequencies in SdH and TEP oscillations are clearly resolved
in addition to relatively low frequencies, i.e, the higher order harmonics and the frequencies $\kappa$, $\xi$, and $\lambda$. However, frequencies of $\alpha$, $\gamma$, and
$\varepsilon$ are only observable for dHvA measurements. Although the frequency $\delta$ = 1.03\,MOe in SdH and TEP spectrum is almost the same as 2$\alpha$ = 0.99\,MOe in dHvA
spectrum, we assigned as different frequency since the first harmonics of this frequency is not seen in the SdH spectrum.

\begin{figure*}
\centering
\includegraphics[width=1\linewidth]{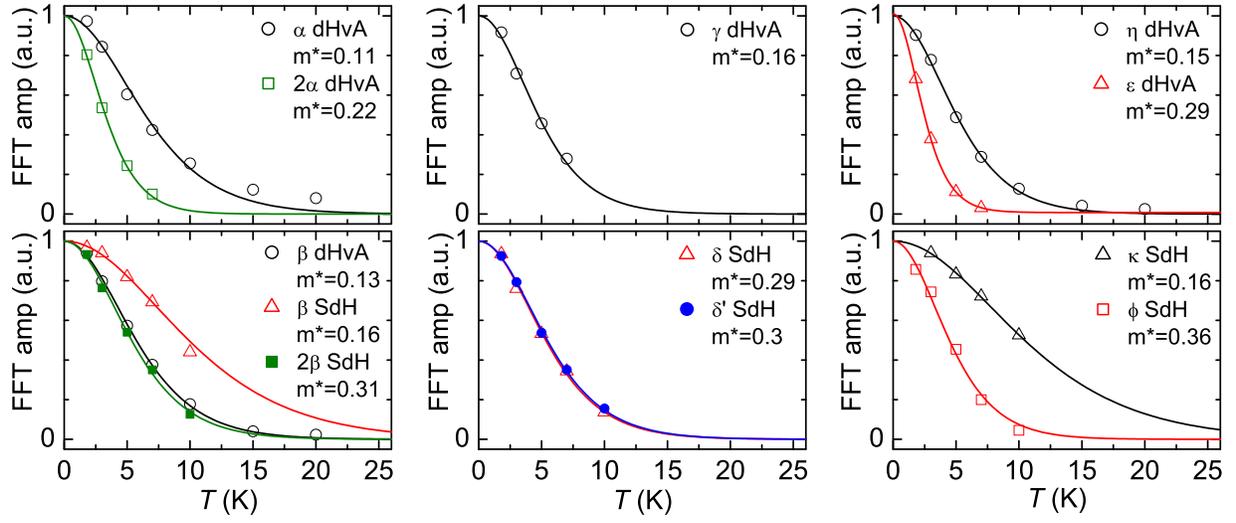}
\caption{Temperature dependence of the dHvA and SdH amplitudes. Solid lines represent the fit to the Lifshitz-Kosevich formula.}
\label{LKfit}%
\end{figure*}

Quantum oscillations are discernible in magnetic fields as low as 5\,kOe for $M(H)$, 35\,kOe for $\rho(H)$, and 20\,kOe for $S(H)$ at the lowest temperature measured; these data imply
very high quality samples as well as very small effective mass of conduction carriers. In order to estimate effective mass of the carriers, dHvA and SdH measurements were carried out
at temperatures from 1.8\,K to greater than 25\,K. The frequencies in FFT spectra do not shift with temperature and most of the first harmonics of the frequencies were clearly
observable as high as 20\,K. The cyclotron effective mass of the carriers from the various orbits were determined by fitting the temperature-dependent amplitude of the oscillations to
the Lifshitz-Kosevich formula \cite{Shoenberg}:
\begin{eqnarray}{\label{LKformula}}
R_{T}=\frac{2\pi{^2}pkT/\beta^{*}H}{\sinh(2\pi{^2}pkT/\beta^{*}H)}\nonumber
\end{eqnarray}
where $\beta^*$ = $e\hbar/m^{*}c$ and the mass $m^*$ is the renormalized, or effective mass. The effective masses, calculated from the temperature dependent amplitude of the
oscillations are shown in Fig. \ref{LKfit}; they range from $m_{\alpha}\sim$ 0.1\,$m_{e}$ to $m_{\xi}\sim$ 0.4\,$m_{e}$, where $m_{e}$ is the bare electron mass. These small effective
masses are consistent with the results of specific heat measurements. The estimated effective masses are summarized in Table \ref{table1}. The effective mass associated with the second
harmonic frequencies, 2$\alpha$ and 2$\beta$, are twice that of first harmonics $\alpha$ and $\beta$, confirming the assumption of these frequencies are higher harmonics. Although the
value of the frequency $\delta$' is integer-multiple of either $\alpha$ or $\delta$, it is not a higher harmonics of either $\alpha$ or $\delta$ because of the inconsistent effective
mass (see Table \ref{table1}). Therefore, $\delta$' is an independent frequency, coming from an external orbit with the area close to 2$\delta$.

\begin{figure}
\begin{center}\leavevmode
\includegraphics[width=1\linewidth]{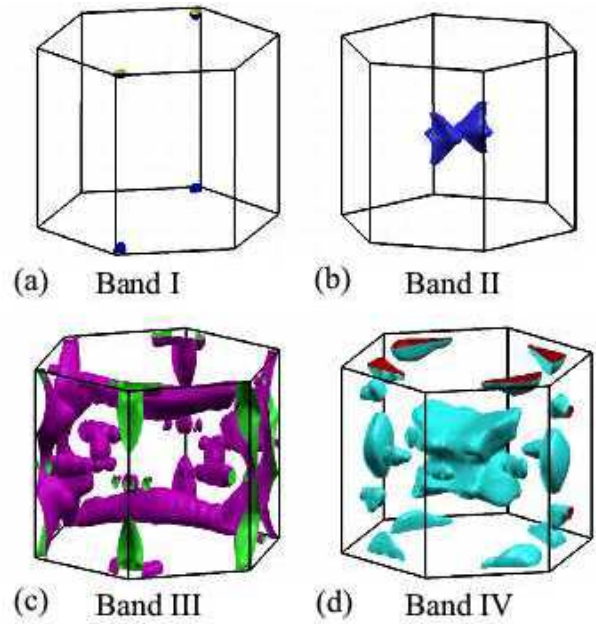}
\end{center} \caption{The theoretically calculated Fermi surface of PtSn$_{4}$.}\label{band}
\end{figure}

As indicated by the dHvA oscillations, for all three orientations, the topology of the Fermi surface of PtSn$_{4}$ is complex. Therefore, band structure calculations have been
performed within the full potential linearized augmented plane wave (FLAPW) method \cite{FLAPW} within local density approximation (LDA) \cite{LDA}. A mesh of 31$\times$31$\times$31
$\vec k$-points was used for Brillouin zone integration and Fermi surface plot, with $R_{MT}*K_{max}$ = 8 , $R_{MT}$ = 2.3 atomic units for both Pt and Sn. We used the room
temperature experimental lattice constants and atomic positions. The original {$\vec k$}-mesh was interpolated onto a mesh ten times as dense. In the calculated total and projected
electronic density of states (DOS), the valence electron bands are derived from hybridized combination of Pt $d$ and Sn $s$ and $p$ states; the electronic states in the energy
interval between -6 and -2 eV are mostly by Pt $d$ states, while the $s$-derived Sn component is distributed mostly below -6 eV. The states at Fermi energy ($E_{F}$) correspond mostly
to $p$ electrons of Sn with a small amount of Pt $d$ states. The $E_{F}$ is located in the local minimum of DOS anomaly. The last fact is usually associated with stability of
particular crystal structure with given concentration of valence electrons. Similar to other semimetals, PtSn$_{4}$ compound has low DOS at $E_{F}$ equals to 2.8 St/eV f.u. and
corresponding $\gamma$ = 4 mJ/mol-K$^{2}$ which is consistent with experimental result. As can be recognized from the results of the calculations (Fig. \ref{band}), the Fermi surface
of PtSn$_{4}$ is three dimensional, and very complex.

From the shape of Fermi surface many branches of frequency can be expected because the Fermi surface has multiple sheets. The relevant branches have not been compared with experimental
frequencies, because there are many possibilities having similar size of extremal cross-sectional areas. The extremal areas in the low frequency region labeled as $\alpha$, $\beta$,
$\gamma$, and $\delta$ can be assigned to band I and the small spherical or ellipsoidal Fermi surface located near to the zone boundary in band III and IV. The higher frequencies near
25\,MOe can be related to two long, cylinderical and cucumber-shaped Fermi surface in band III or a large pillow-shaped Fermi surface in band IV. It would be necessary to measure
frequencies as a function of angle between the crystallographic axes in order to make estimates of the Fermi surface topology.

\section{Summary}
High quality single crystals of PtSn$_{4}$ were grown from high temperature solution. Based on measurements of temperature and field dependencies of thermodynamic and transport
properties, PtSn$_{4}$ can be classified as a metallic system with an axial magnetic and electronic anisotropy. In zero field the low residual resistivity and large resistivity ratio
between 2 and 300\,K indicate a high quality of PtSn$_{4}$ samples. The high quality of samples is further supported by the observation of very large MR and pronounced SdH and dHvA
oscillations. The Hall resistivity and thermoelectric power measurement indicate a sign reversal below 50 K, caused by the dramatic change in the carrier density and mobility of the
electron band. The application of a magnetic field causes not only a large magentoresistance but also a large magneto-thermoelectric power effect. These large magnetic field effects
on transport properties are mainly due to different mobility changes for the different types of carriers. It is found that the Kohler's rule is valid in the transverse
magnetoresistance of PtSn$_{4}$, scaling with $\Delta\rho/\rho(0)\simeq H^{1.8}$. Finally, the observed quantum oscillations and an aniostropic field response on MR indicate a very
complicate electronic structure of this compound.

\begin{acknowledgments}
Work at the Ames Laboratory was supported by the U.S. Department of Energy Basic Energy Sciences under Contract No. DE-AC02-07CH11358.
\end{acknowledgments}

\begin{table*}
\caption{Frequencies and effective masses $m^{*}$ obtained from the dHvA ($M(H)$), SdH ($\rho(H)$) and TEP ($S(H)$) oscillations. $m_{e}$ is the
bare electron mass.}
\label{table1}%
\begin{ruledtabular}
\begin{tabular}{cccccc|cc|cc}
\multicolumn{6}{c}{\textbf{H}\,$\parallel$\,\textbf{b}} &\multicolumn{2}{l}{\textbf{H}\,$\perp$\,\textbf{b}I} &\multicolumn{2}{l}{\textbf{H}\,$\perp$\,\textbf{b}II}    \\
                &\multicolumn{2}{c}{dHvA}   &\multicolumn{2}{c}{SdH}& $S(H)$  &              & dHvA &              & dHvA   \\
                &   (MOe)   & $m^{*}/m_{e}$ & (MOe)  & $m^{*}/m_{e}$& (MOe)   &              &(MOe) &              & (MOe)  \\ \hline
$\alpha$        &   0.52    & 0.11          &        &              &         & $\alpha_{1}$ & 0.52 & $\alpha_{2}$ & 0.57   \\
$\beta$         &   0.63    & 0.13          & 0.61   &   0.16       & 0.61    & $\beta_{1}$  & 0.73 & $\beta_{2}$  & 0.68   \\
$\gamma$        &   0.84    & 0.16          &        &              &         & $2\alpha_{1}$& 1.04 & $2\alpha_{2}$& 1.10   \\
$2\alpha$       &   0.99    & 0.23          &        &              &         & $2\beta_{1}$ & 1.41 & $2\beta_{2}$ & 1.36   \\
$\delta$        &           &               & 1.03   &   0.29       & 1.03    & $\gamma_{1}$ & 2.46 &              &        \\
$2\beta$        &   1.26    &               & 1.26   &   0.31       & 1.27    & $\delta_{1}$ & 2.62 & $\delta_{2}$ & 2.57   \\
$\eta$          &   1.78    & 0.15          & 1.79   &              & 1.81    & $2\gamma_{1}$& 4.93 &              &       \\
$\delta$'       &   1.99    &               & 1.98   &   0.30       & 2.07    & $2\delta_{1}$& 5.71 &              &       \\
$\varepsilon$   &   3.25    & 0.29          &        &              &         & $\lambda_{1}$& 7.91 &              &       \\
$2\eta$         &   3.56    &               &        &              &         & $\psi_{1}$   & 9.33 &              &       \\
$\delta$"       &   4.03    &               & 4.08   &              & 4.07    & $\nu_{1}$    &      &              &       \\
$2\varepsilon$  &   6.45    &               &        &              &         &              &      &              &       \\
$\kappa$        &           &               & 3.02   &   0.16       & 3.07    &              &      &              &       \\
$\xi$           &           &               & 3.44   &   0.36       & 3.42    &              &      &              &       \\
$\lambda$       &           &               & 5.19   &              & 5.18    &              &      &              &       \\
$2\kappa$       &           &               & 6.11   &              & 6.14    &              &      &              &       \\
$2\xi$          &           &               & 6.87   &              & 6.83    &              &      &              &       \\
\end{tabular}
\end{ruledtabular}
\end{table*}

\clearpage

\end{document}